# Water ice in the debris disk around HD 181327

Chen Xie[1,†], Christine Chen[1,2], Carey M. Lisse[3], Dean C. Hines[2], Tracy Beck[2], Sarah K. Betti[2], Noemí Pinilla-Alonso[4,5], Carl Ingebretsen[1], Kadin Worthen[1], András Gáspár[6], Schuyler G. Wolff[6], Bryce T. Bolin[7], Laurent Pueyo[2], Marshall D. Perrin[2], John A. Stansberry[2], and Jarron M. Leisenring[6]

1. William H. Miller III Department of Physics and Astronomy, John's Hopkins University, 3400 N. Charles Street, Baltimore, MD 21218, USA
2. Space Telescope Science Institute, 3700 San Martin Drive, Baltimore, MD 21218, USA
3. Johns Hopkins University Applied Physics Laboratory, 11100 Johns Hopkins Road, Laurel, MD 20723, USA
4. Institute of Space Sciences and Technologies of Asturias, University of Oviedo, Spain
5. Department of Physics, University of Central Florida, Orlando, FL, USA
6. Steward Observatory and the Department of Astronomy, The University of Arizona Tucson, 933 N. Cherry Avenue, Tucson, AZ 85721, USA
7. Goddard Space Flight Center, 8800 Greenbelt Road, Greenbelt, MD 20771, USA
† e-mail: cxie21@jh.edu

## Summary

Debris disks are exoplanetary systems that contain planets, minor bodies (i.e., asteroids, Kuiper belt objects, comets, etc.), and micron-sized debris dust[1]. Since water ice is the most common frozen volatile, it plays an essential role in the formation of planets[2,3] and minor bodies. Although water ice has been commonly found in Kuiper belt objects and comets in the Solar System[4], no definitive evidence for water ice in debris disks has been obtained to date[1]. Here, we report the discovery of water ice in the HD 181327 debris disk using the James Webb Space Telescope Near-Infrared Spectrograph. We detect the solid-state broad absorption feature of water ice at 3 μm including a distinct Fresnel peak feature at 3.1 μm, indicative of large, crystalline water-ice particles. Gradients in the water-ice feature as a function of stellocentric distance reveal a dynamic environment in which water ice is destroyed and replenished. We estimate water-ice mass fractions ranging from 0.1% at ~85 au to 21% at ~113 au, indicating the presence of a water-ice reservoir in the HD 181327 disk beyond the snowline. The icy bodies that release water ice in HD 181327 are likely the extra-solar counterparts of water-ice-rich Kuiper belt objects in our Solar System.

## Main text

We observed HD 181327 using the James Webb Space Telescope (JWST) Near-Infrared Spectrograph (NIRSpec)[5,6] integral field unit (IFU; spectral resolving power of approximately 100) as part of the General Observer program 1563 (PI: C. Chen; Methods). HD 181327 is an F6-type[7] star at 47.72 ± 0.05 pc[8], with an estimated age of ~18.5 Myr[9]. The ring-like disk around HD 181327 has a brightness peak at 84 au with a width of ~25 au and an extended disk halo[10–12]. We spatially resolve the disk after removing the stellar point spread function (PSF), as shown in Fig. 1. The estimated temperature of black bodies in the HD 181327 debris ring is ~40 K[11], consistent with the presence of planetesimals beyond the water ice snowline. Moreover, Atacama Large Millimeter/submillimeter Array (ALMA) observations have detected CO gas[12] believed to be liberated in collisions among volatile-rich planetesimals[13] similar to comets and Kuiper belt objects (KBOs) in our Solar System. As a result, the HD 181327 debris disk is believed to be a young, extra solar analog to the Solar System's Kuiper belt making it an ideal target for water ice searches. We extract disk reflectance spectra at a scattering angle of ~90º and stellocentric distances between 80–120 au as detailed in Methods and Fig. 1, showing the scattering efficiency of dust particles in the disk.

The disk reflectance spectrum at 90-105 au has a broad bowl-shaped dip between 2.7 and 3.4 μm (Fig. 1), consistent with the 3 μm feature of water ice ($H_2O$). Within the solid-state feature at 3 μm, we detect a narrow and strong peak at 3.1 μm in both the disk spectrum and its reflectance spectrum (Extended Data Fig. 3), which we attribute to the Fresnel peak of crystalline water ice, as observed in the spectra of Saturn's rings[14] and KBOs[15–17]. The Fresnel peak is generated by refractive lensing of large (i.e., ~1 mm), crystalline water ice particles[18,19]. No significant difference is found in the normalized reflectance spectra extracted separately from the two sides of the disk (Extended Data Fig. 5). Since the bowl-shaped dip and the Fresnel peak are not artifacts generated by our post-processing or our flux corrections (Methods and Extended Data Figs. 1–5), this is an unambiguous detection of water ice in a debris disk[1], confirming the previous hint of water ice in the observed spectral energy distribution (SED) of HD 181327[11,20]. We also find a tentative detection (2.3 $\sigma$) of $CO_2$-ice absorption at 4.25 μm in the spectrum at 105-120 au (Fig. 1). The detections of water ice, CO gas[12], and possible $CO_2$ ice indicate the presence of a volatile-rich reservoir in the disk around HD 181327, rich in materials found in icy bodies in the outer parts of our Solar System.

The disk reflectance spectra show clear gradients in the water-ice band depth and the Fresnel peak at different stellocentric distances (i.e., ~85, ~97, and ~113 au). The spectra reveal dust with varying composition and grain size as a function of stellocentric distance (Table 1), consistent with the lack of water ice at 80–90 au and a mixture of water ice and other dust species at larger distances (90–120 au). The modeling results suggest the water-ice mass fractions are about 0.1%, 7.5%, and 21.0% at 80–90 au, 90–105 au, and 105–120 au, respectively. The spectrum at 105–120 au also shows the 4.5 μm water-ice feature, consistent with more water-ice particles at larger distances. The disappearance of the Fresnel peak at >105 au is coincident with the outer edge (~100 au) of the planetesimal belt observed by ALMA[12]. It is unlikely that the distribution of water ice is intrinsically different within a relatively narrow region of 10 au. Therefore, the gradient suggests that the HD 181327 disk is dynamic, with micron-sized icy particles being continuously created and destroyed in the planetesimal belt beyond the snowline.

We hypothesize that the water-ice gradient may be created by photodesorption of water ice at the inner part of the disk. We calculate the photodesorption timescale of water ice, as detailed in Methods and Fig. 2. Photodesorption can efficiently remove water ice from icy grains that are smaller than 10 μm (see also Extended Data Fig. 7). Since the blowout size of the HD 181327 grains is ~1 μm (Methods), photodesorption is expected to remove water ice from the smallest dust grains in the distribution. The lack of water ice in the innermost region is consistent with photodesorption of water ice at 80–90 au. The presence of water ice at 90–105 au can be explained if the replenishment of surface water ice is faster than the erosion rate. If the disk is marginally optically thick at ultraviolet (UV) wavelengths in the radial direction of the disk, then the dust may shield icy particles from photodesorption. The optical depth required depends on the replenishment rate of water ice. For example, we find that a water-ice replenishing timescale of $2 \times 10^4$ yr and an optical depth of 3 between 70 and 120 au can explain the observed radial gradient of water ice (Fig. 2 and Methods). Thermal sublimation is responsible for the narrow (8 au) devolatilized ring of material (100 K) at ~75 au[21] around early

A-star HR 4796A and the turn-on of cometary activity in the comets of the Solar System. We exclude this mechanism due to the low dust temperature of <60 K (Extended Data Fig. 6) and the exponential temperature dependence of thermal sublimation[22,23]. Since water ice is detected and the estimated photodesorption timescale of water ice grains with a radius of 10 μm is significantly shorter than the system age of ~18.5 Myr[9], surface water ice must be replenished through collisional scouring of old icy bodies and/or continual production of new icy dust by collisions of icy parent bodies[11].

We find that iron sulfide (FeS) and olivine ($MgFeSiO_4$) can reproduce the overall shape of the continuum, as detailed in Methods and Table 1. Instead of producing distinct line-like features, FeS and olivine contribute low-frequency structure to the overall slope of the disk reflectance spectrum. The presence of FeS in the HD 181327 debris disk has been previously inferred from polarized and total intensity scattered light observations at the H band (1.6 μm)[24]. FeS has been found in micrometeorite and comet samples[25], in the ejecta of the Deep Impact comet excavation experiment[26], and used to reproduce observed photo-polarimetric properties of the HR 4796 disk at near-infrared wavelengths[27]. Olivine is a very common refractory dust species found in protoplanetary disks, debris disks[28], and Solar System comets and asteroids[29]. The HD 181327 Spitzer SED has previously been fit using olivine dust grains[11,20]. For our modeling, we note that olivine and pyroxene have similar scattered light spectra at 1–5 μm, making fits of the scattered light spectra degenerate. Although olivine can be replaced with pyroxene and still produce good fits (i.e., 80–90 au), we used only olivine in the dust model for simplicity. Observing unique solid-state features is necessary to unambiguously identify the composition of both icy and refractory grains.

Simultaneous modeling ($\chi_v^2$<2) of the water-ice feature and the continuum supports the detection of water ice and provides detailed constraints on the grain properties. We find that a model with two dust populations can reproduce the observed water-ice feature at 3 μm while the single population model failed (Extended Data Fig. 3 and Methods). The two-population model (with dust properties listed in Table 1) includes two separate dust mediums while the single-population model includes a well-mixed dust medium. The presence of two dust populations within the best-fit model suggest that the dust grains are probably derived from two distinct reservoirs, either totally separate or loosely mixed together. Interestingly, our model is consistent with having no significant amount ($V_{\mathrm{olivine}}$<5%) of olivine mixed with water ice. Nearly pure water ice has been observed towards the surface of Haumea family objects and is believed to be the result of collisional resurfacing of differentiated bodies[30–32]. Besides water ice, the spectra at shorter wavelengths (<2.5 μm) become bluer at larger distances, that is consistent with smaller dust particles at larger distances, similar to those found in visual observations[33]. Our modeling also finds the minimum grain size decreases with increasing distance as $a_{\mathrm{min}}$= 1.6, 1.4, and 0.8 μm at 85, 97, and 113 au, respectively. Together with the size distributions ($p$) ranging from –3.3 to –4.0, the micron-sized dust grains follow the collisional nature of the main ring with a halo of small particles blown out by radiation pressure[24,33].

The disk reflectance spectrum shows no significant water-ice features at 1.5 and 2 μm or CO features at ~4.7 μm (Fig. 1). The lack of 1.5 and 2 μm features could be attributed to small particle size[34] (Table 1). For small particles, the 1.5 and 2 μm features are much weaker than the 3 μm feature, making them more challenging to detect. Although CO gas is expected to be released from icy bodies[13] or produced by $CO_2$ ice photodesorption[12], we do not detect significant CO gas or ice features at ~4.7 μm, possibly due to the relatively large uncertainty at longer wavelengths. Furthermore, the limited spectral signal-to-noise ratio makes it difficult to discern faint spectral features of additional molecules, such as methanol and ammonia. Model fitting is compatible with no significant ($V_{\mathrm{carbon}}$<5%) amorphous carbon in the spectrum. However, due to the degeneracy of the flat and featureless spectrum of amorphous carbon, it is hard to estimate its abundance (Methods).

The HD 181327 disk is composed of dust particles generated by collisions among planetesimals beyond the snowline[1]. The composition of the dust is expected to reflect that of the parent planetesimals if the dust remains unaltered as it was released. Hence, comparing the spectral feature of the dust species in disks and Solar System objects may help us better understand the origin of the dust in the disk beyond the snowline. To better compare the water-ice features, we subtracted the disk continuum because the spectral shape of water ice was affected by the underlying continuum (Methods). Among spectra of minor bodies in our Solar System[15,16,34,35], we find some of the water-ice-rich KBOs[17,36] have the closest spectra of the water ice feature at 3 μm to the spectra of the young disk (~18.5 Myr) around HD 181327 (Fig. 3). Water-ice-rich KBOs have a deeper band depth of the water-ice feature at 3 μm than that of the disk, which are probably the result of the larger average water-ice grain sizes and different water-ice abundance on the KBO surfaces. Since the icy grains in the HD 181327 disk are thought to be replenished through collisions among icy parent bodies[11], these icy bodies could likely be the extra-solar counterparts of water-ice-rich KBOs in our Solar System.

Comparing and contrasting the small dust grains in the debris disk and minor bodies in our Solar System can provide insight into the formation of KBOs and the retention and replenishment of volatiles inside and outside of the Solar System. For example, there is still much ongoing debate as to when KBO objects like Pluto formed[38,39] and when Kuiper belt self-stirring becomes important[40]. With HD 181327, at least, we have an example of a system undergoing active collisions in its planetesimal belt between 9–18 Myr after its nominal disk clearing time at <10 Myr[41] age, suggesting a rapid formation of the largest KBOs in this system. As such, it also provides insights into the physics of dust scattering properties at different scales and environments. JWST NIRSpec is opening a new window on the study of volatile ices in debris disks by detecting solid-state features from water ice, indicating the presence of water ice and other volatiles in a CO-rich planetesimal belt beyond the snowline.

**Table 1 | Best-fit dust grain parameters at different distances**

| Dust properties | Middle ring | Outer ring | Halo |
|---|---|---|---|
| Distance (au) | 80–90 | 90–105 | 105–120 |
| Model ID | #1 | #2 | #3 |
| Dust population #1 | | | |
| $f_{\mathrm{pop1}}$ | 1 | $0.46^{+0.06}_{-0.05}$ | $0.60^{+0.04}_{-0.04}$ |
| $a_{\mathrm{min}}$ (μm) | $1.61^{+0.04}_{-0.03}$ | $1.39^{+0.03}_{-0.02}$ | $0.75^{+0.02}_{-0.02}$ |
| $a_{\mathrm{max}}$ (μm) | $394^{+334}_{-168}$ | $537^{+259}_{-265}$ | $498^{+334}_{-323}$ |
| $p$ | $-3.26^{+0.05}_{-0.03}$ | $-4.04^{+0.11}_{-0.12}$ | $-3.71^{+0.08}_{-0.09}$ |
| $V_{\mathrm{porosity}}$ | $0.02^{+0.02}_{-0.03}$ | $0.04^{+0.04}_{-0.07}$ | $0.59^{+0.04}_{-0.02}$ |
| $V_{\mathrm{H2O;\ amorphous}}$ | $0.004^{+0.001}_{-0.001}$ | $0.002^{+0.002}_{-0.002}$ | $0.017^{+0.006}_{-0.009}$ |
| $V_{\mathrm{H2O;\ crystalline}}$ | $0.001^{+0.001}_{-0.001}$ | $0.001^{+0.001}_{-0.001}$ | $0.001^{+0.001}_{-0.001}$ |
| $V_{\mathrm{FeS}}$ | $0.04^{+0.01}_{-0.01}$ | $0.01^{+0.02}_{-0.01}$ | $0.37^{+0.01}_{-0.01}$ |
| $V_{\mathrm{olivine}}$ | $0.94^{+0.02}_{-0.03}$ | $0.95^{+0.03}_{-0.07}$ | $0.02^{+0.04}_{-0.02}$ |
| Dust population #2 | | | |
| $f_{\mathrm{pop2}}$ | -- | $0.54^{+0.06}_{-0.05}$ | $0.40^{+0.04}_{-0.04}$ |
| $a_{\mathrm{min}}$ (μm) | -- | $3.08^{+0.26}_{-0.26}$ | $0.90^{+0.14}_{-0.12}$ |
| $a_{\mathrm{max}}$ (μm) | -- | $509^{+283}_{-252}$ | $542^{+314}_{-296}$ |
| $p$ | -- | $-4.01^{+0.09}_{-0.09}$ | $-3.71^{+0.09}_{-0.10}$ |
| $V_{\mathrm{porosity}}$ | -- | $0.72^{+0.03}_{-0.03}$ | $0.19^{+0.11}_{-0.13}$ |
| $V_{\mathrm{H2O;\ crystalline}}$ | -- | $0.28^{+0.03}_{-0.03}$ | $0.80^{+0.11}_{-0.13}$ |
| $V_{\mathrm{olivine}}$ | -- | $0.002^{+0.005}_{-0.001}$ | $0.01^{+0.01}_{-0.01}$ |
| $\chi^2_v$ | 1.56 | 1.62 | 1.61 |

Error bars represent 68% uncertainty. The corresponding corner plots are presented in Extended Data Figs. 9–11. The volume fractions of $H_2O$ in dust population #1 are close to zero, supporting the use of two dust populations, with a water-ice-depleted population and a water-ice-rich population. Rows: (1) stellocentric distance; (4) first dust population fraction ($f_{\mathrm{pop1}}$); (5–6) minimum and maximum grain radius ($a_{\mathrm{min}}$ and $a_{\mathrm{max}}$); (7) grain size distribution ($a^p$); (8–12) volume fractions ($V$) of porosity, amorphous water ice, crystalline water ice, iron sulfide, and olivine; (14) second dust population fraction ($f_{\mathrm{pop2}}$=1–$f_{\mathrm{pop1}}$); (21) total reduced $\chi^2$.

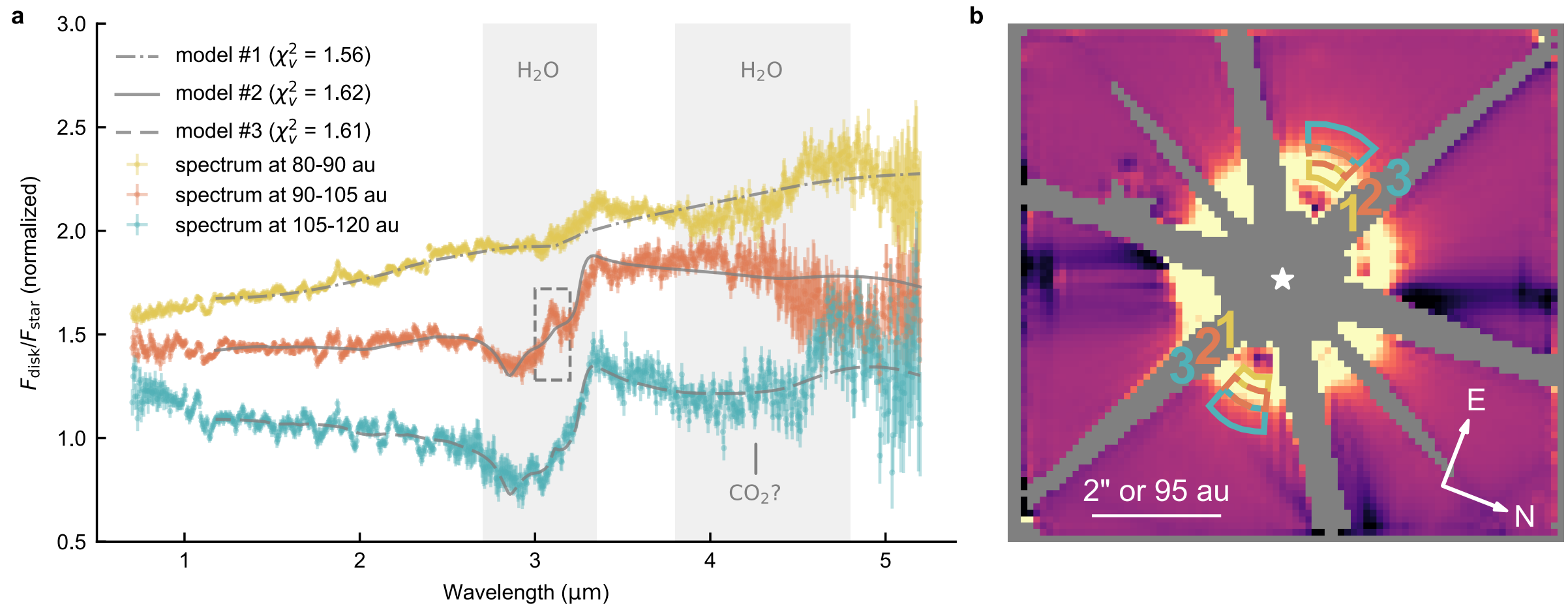

**Fig. 1 | Disk reflectance spectra at different stellocentric distances. a**, The disk reflectance spectra at 80–90 au (region 1, yellow), 90–105 au (region 2, orange), and 105–120 au (region 3, cyan), corresponding to the locations shown in panel b. The disk reflectance spectrum extracted from 80–90 au lacks water ice features, while that at 90–105 au shows the clear presence of water ice at ~3 μm and a Fresnel peak at 3.1 μm (dashed box), and that at 105–120 au shows a deeper water-ice feature at 3 μm and the water-ice feature at ~4.5 μm. The best-fit dust models are shown in gray. The corresponding model parameters are listed in Table 1. Error bars represent 1 s.d. **b**, The 0.6–5.2 μm image of the HD 181327 debris disk after removal of the stellar PSF. The apertures used to extract the spectra in panel a are marked as 1, 2, and 3 using corresponding colors. The location of the host star is marked with the white star. Regions with large PSF subtraction residuals (e.g., diffraction spikes and the central region**)** are masked in gray. The dark horizontal strip in the image is an artifact caused by detector saturation.

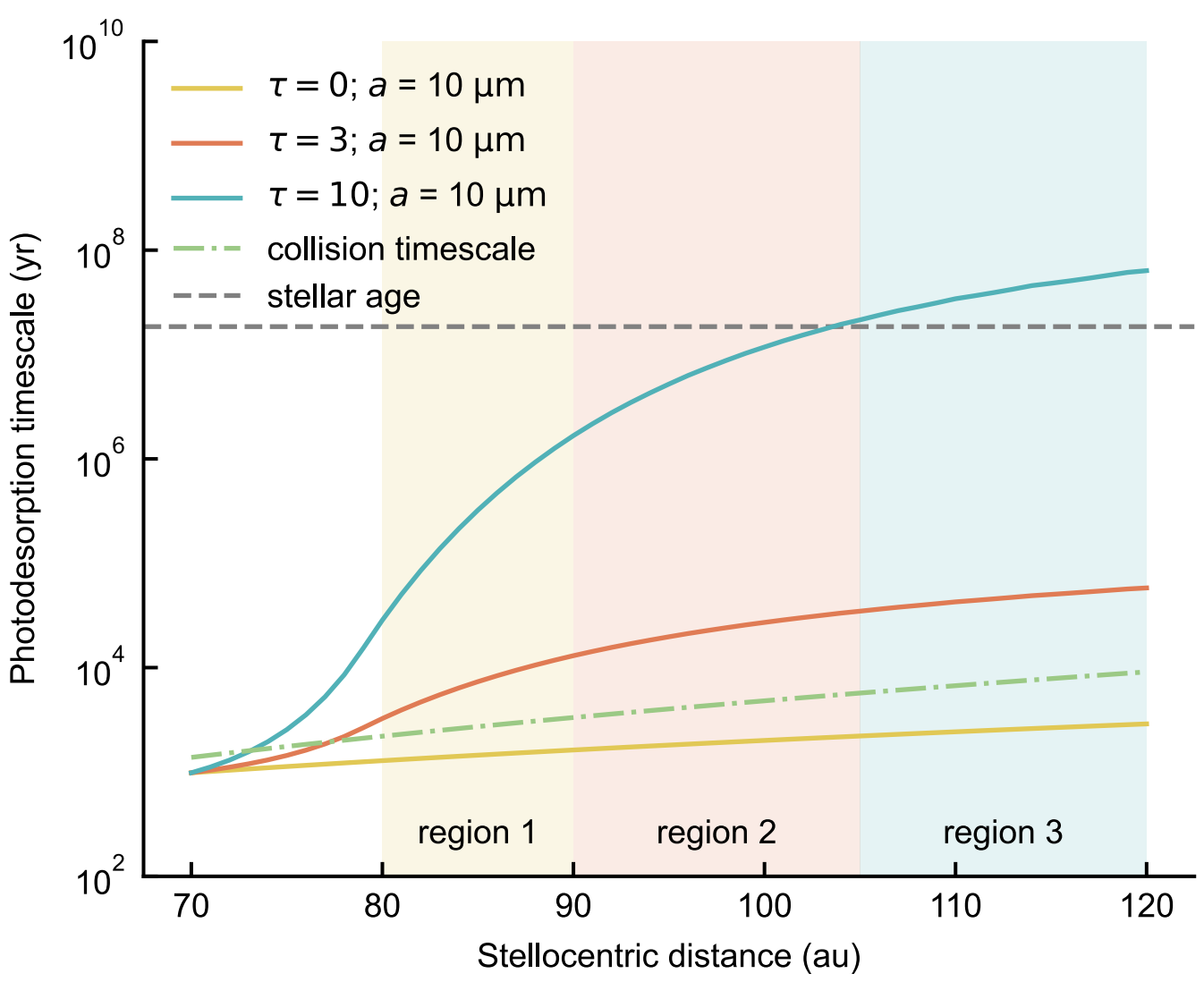

**Fig. 2 | Photodesorption timescales of water ice as a function of stellocentric distance.** The photodesorption timescale for destroying water ice from an icy grain of 10 μm radius as a function of distance, assuming different optical depths ($\tau$) in the radial direction of the disk between 70 and 120 au. Assuming a constant erosion rate over time, the photodesorption timescale is linearly proportional to the grain size, with larger grains taking more time to erode (Methods). Color-shaded regions indicate the distances corresponding to the spectral extracting regions in Fig. 1. The collision timescale of dust grains sets the highest replenishment rate of water ice if fragments produced by the collision always contain water ice on the surface. The

stellar age of 18.5 Myr sets the lowest replenishment rate of water ice. The detection of water ice indicates the photodesorption timescale is longer than the actual replenishment rate of water ice.

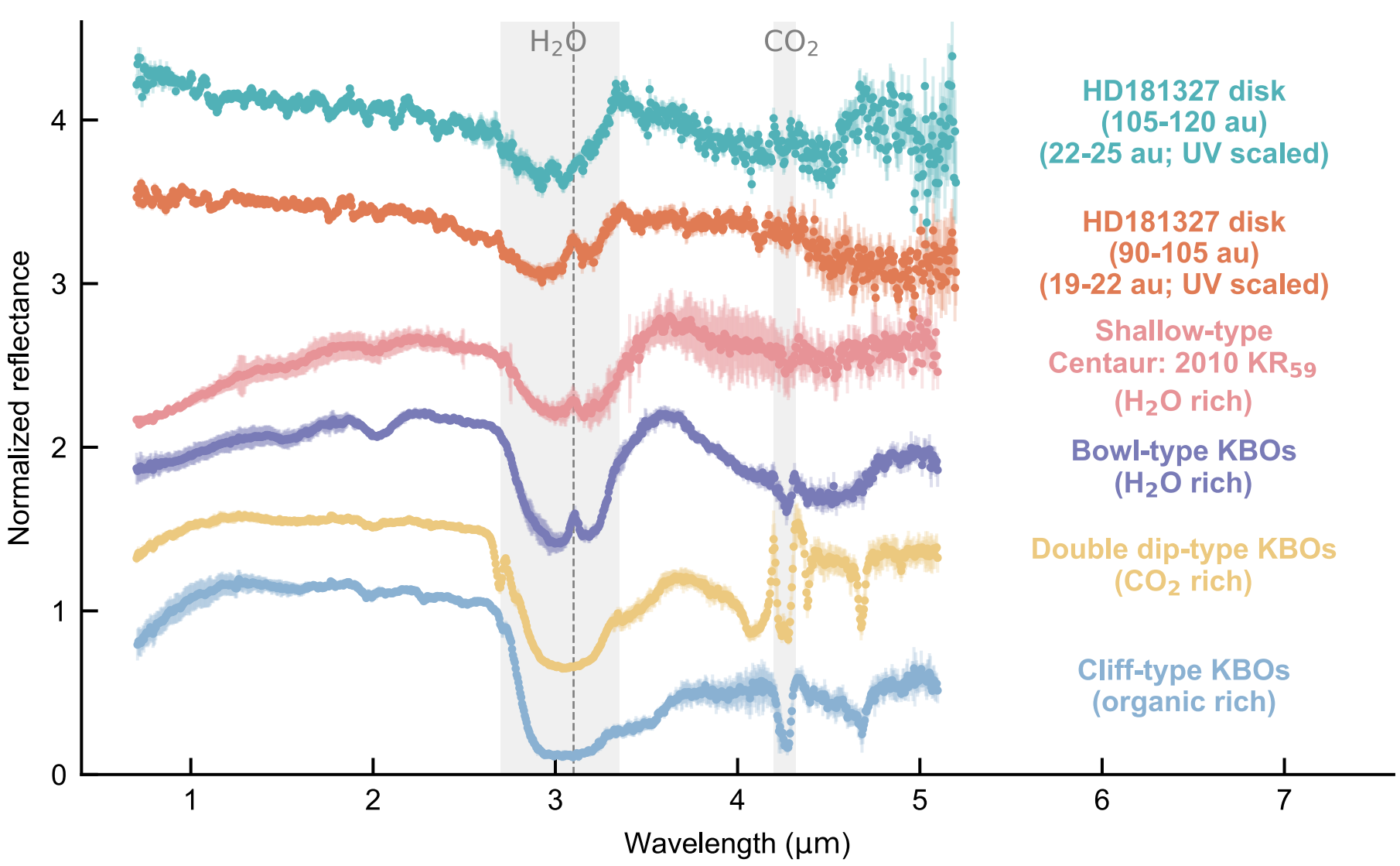

**Fig. 3 | Spectral comparison between the HD 181327 disk and icy bodies in the Solar System.** By subtracting the disk continuum (Methods and Extended Data Fig. 8), we obtain the water-ice-dominated disk spectra at 90–105 au and 105–120 au, showing the water-ice feature at 3 μm. Error bars represent 1 s.d. The disk distance is scaled to match the solar UV luminosity (Methods). Four types of icy body spectra are also shown for comparison, adopted from refs. [17,36], with the principal ice included in the label. The HD 181327 spectrum at 105–120 au shows a 4.25 μm feature that may be a $CO_2$ ice feature (gray shaded); CO2 is commonly observed in icy KBO spectra[37]. Although the band depths of water ice are different potentially caused by different grain sizes and water-ice abundance, the spectra of water-ice-rich KBOs and the disk show a similarity in water ice, considering the fact that the two spectra came from different sources (i.e., micron-sized dust grains in a debris disk beyond the snowline and the surface regolith of icy bodies in the Solar System). The location of the Fresnel peak at 3.1μm is marked by the vertical dashed line.

# Methods

### HD 181327 system

HD 181327 is an F6-type[7] star member of the β Pictoris moving group at 47.72 ± 0.05 pc[8], with an estimated age of ~18.5 Myr[9]. The debris disk around HD 181327 was discovered using IRAS because the star possesses IR excess emission at 25 and 60 μm[42]. The Hubble Space Telescope (HST) spatially resolved the debris disk via coronagraphic imaging[10,43], showing a ~25 au wide ring with a brightness peak at ~84 au[24] and an inclination of 30.2°[24]. The ring has a clear inner edge at ~76 au[24] and an extended halo (up to ~400 au) detected in the visual image[33]. Moreover, the width of the forward-scattering peak increases with stellocentric distances, suggesting smaller dust particles at larger distances[33]. ALMA observations revealed the planetesimal belt at 1.3 mm and found the semi-major axis of the ring to be smaller by ~4 au as determined from visual observations[12,24]. This is consistent with a collisional and large-particle-dominated main ring producing a halo of secondary small particles blowing out by radiation pressure[24].

Besides the detection of dust continuum, ALMA also detected CO emission[12] in the HD 181327 disk. The detection of CO emission is consistent with the presence of an underlying ice reservoir that also includes CO[13]. Thus, the observed CO emission in HD 181327 is consistent with the presence of volatile-rich planetesimals[12]. Near-infrared polarimetry of the Solar System's comets and the HD 181327 disk also show a striking similarity[24]. Furthermore, SED fitting suggested that the peak at 60–75 μm was too narrow to be fitted by a single-temperature blackbody and could be produced by emission from crystalline water ice[11]. The hint of water ice and CO emission in the HD 181327 system, like the small icy bodies found in our own Kuiper belt, encouraged the search for water ice. Hence, we carried out a detailed reflectance spectroscopic study of the debris disk around HD 181327 using the JWST/NIRSpec IFU to spatially resolve and extract its near-infrared spectra, covering the solid-state features of water-ice at 1.5, 2.0, and 3.0 μm.

**Observations and data reduction**

The HD 181327 system was observed with the JWST NIRSpec[5,6] IFU[44] on 6 September 2023 as part of the General Observer program 1563 (PI: C. Chen), using the PRISM/CLEAR configuration. The corresponding PSF calibrator (Iota Mic) was observed in a non-interruptible sequence. The disk was observed in NRSIRS2RAPID read-out mode with a nine-point dither pattern for a total on-source exposure time of 5,256 s. Eight mosaic tiles were used to cover the entire disk, while the star was excluded from observations as it would have saturated a large portion of the field. Both the science star and the PSF calibrator were observed by using the same observing settings. The mosaic IFU observations produce a three-dimensional spectral cube in a final field of view of 8.2" × 7.8" with a pixel size of 0.1" × 0.1", covering 0.6–5.3 μm with a spectral resolution of ~100. At the distance of HD 181327, the full width at half maximum (FWHM) of the NIRspec PSF is ~4.8 au at 3 μm.

We processed the HD 181327 and Iota Mic data using the standard JWST Science Calibration pipeline[45] (v1.13.3) stages 1 to 3. The Calibration Reference Data System of jwst_1180.pmap was used as the calibration files for bias subtraction, flat fielding, wavelength calibration, and flux calibration. Specifically, the data reduction consists of three main steps. First, the Detector1Pipeline.call of the JWST pipeline was used to process the uncalibrated raw data with the default parameters. The Spec2Pipeline.call was then used with default parameters, with background subtraction and extracting 1D spectrum skipped. The output cube (*rate.fits) is in the detector coordinates ('ifualign' frame). Finally, the Spec3Pipeline.call was used with default parameters, with the output cube in the detector coordinates to avoid additional noise introduced by interpolation. Both HD 181327 and Iota Mic were processed with the same parameters.

We recentered the HD 181327 and Iota Mic data cubes by using their diffraction spikes to determine their star position. To increase the accuracy for the measurement of the star position, we median combined the first 500 spectral channels to boost the signal-to-noise ratio and applied a 25 × 25 pixels median filter to enhance the spike structures. We didn't include the channels at longer wavelengths because the diffraction spikes are fainter and wider at longer wavelengths. We next applied a $r_{\text{sep}}^2$ correction to equally weight the pixels of the spikes at different angular separations ($r_{\text{sep}}$) from the PSF center. After that, we determined the PSF center for a given data cube by interactively using radonCenter[46], which is a Radon Transform-based method[47] for measuring the PSF center. Such a method has been a common approach to determining the star position of HST observations. Specifically, radonCenter performs line integrals in the azimuthal direction for different positions in the search region (± 5 pixels). The sum of the line integral of each diffraction spike reaches its maximum value at the star position. All eight diffraction spikes in NIRSpec have been used to accurately constrain the star position, reaching the precision of <0.1 pixels (<10 mas). After the HD 181327 and Iota Mic data cubes were aligned, we explored their offset of the position angles using diffraction spikes. In the detector frame, we found that the difference in the position angle between our data cubes was less than 0.02 degrees. Hence, no angular alignment was performed.

**Reference-star differential imaging (RDI)**

The NIRSpec IFU does not have a coronagraph to suppress the light from the host star. As a result, PSF subtraction is essential to detect the circumstellar disk. We performed PSF subtraction from every spectral channel using reference-star differential imaging (RDI)[48,49] to remove the stellar PSF in the spatial direction before extracting the disk spectrum. Iota Mic was observed with the same observing strategies as HD 181327 and does not have other sources (stars, companions, or a disk) in the field of view. Thus, the observation of Iota Mic provides an empirical measurement of the PSF that can be used in classical PSF subtraction. The PSF subtraction process was performed in two steps: (1) making masks and (2) finding the scaling factor between the science target and the PSF reference to perform RDI in each spectral channel.

We created a total mask to increase the accuracy of fitting the scaling factor in the second step. The total mask consists of four separate masks: (1) an inner mask, (2) an outer mask, (3) a spikes mask, and (4) a disk mask. Since HD 181327 (K-band: 5.9 mag[50]) and Iota Mic (K-band: 4.3 mag[50]) have high apparent brightness, they saturated at the angular offsets of approximately 0.7" and 1.4" at 1.13 μm for HD 181327 and Iota Mic, respectively. To avoid the saturated region, we adopted an inner mask with a radius of 15 pixels (or 1.5"). To avoid the influence of noisy pixels at the edge of the field of view, we adopted an outer mask with a radius of 38 pixels (or 3.8"). The eight diffraction spikes are prominent features and their fluxes do not match the corresponding stellar photosphere model. This is because the optical effect from the varying PSF width as a function of wavelength, variation in speckles, and structure in the diffraction spikes themselves cause flux variation that is higher than the flux in the PSF diffraction spike at >1.5". Hence the diffraction spikes cannot be used to find the scaling factor to match the science image with the PSF reference. Therefore, we also masked the eight diffraction spikes. Finally, to avoid the oversubtraction effect[51], we created a disk mask to avoid including disk flux when fitting the scaling factor in RDI. The oversubtraction effect is severe if the disk has a compatible brightness with the PSF at the disk locations. After applying four masks, we used the remaining annular regions to calculate the scaling factor by matching the science and reference images in each spectral channel.

We calculated the scaling factor for each spectral channel by minimizing the cost function ($\mathsf{C}$) as:

$$\arg\min_{\lambda,i} \mathsf{C} \;=\; \log\left(\sum_{i=1}^{N_{\text{pix}}}\left((I_i - f_{\text{RDI}}R_i)M_i\right)^2\right),$$

where $f_{\text{RDI}}$ is the scaling factor in RDI, $\lambda$ is the wavelength, $i$ is the image pixel index, $N_{\text{pix}}$ is the total number of pixels, $I$ represents the science image, $R$ is the PSF reference image, and $M$ is the image mask. Thus, the scale factor ($f_{\text{RDI}}$) matches the halo of the stellar PSF in the science and the reference data for each wavelength. We then scaled the PSF reference image to subtract the stellar PSF from the science image for each spectral channel to reveal the debris disk. We present the obtained scaling factor ($f_{\rm RDI}$) in Extended Data Fig. 1, showing no correlations with the observed water-ice features. The limited variation in scaling factors between 1.5–5 μm suggests a good match between the science and reference stars. The small fluctuation in the scaling factor could be caused by the spectral difference between science and reference stars, bad/hot pixels in the fitting region, and instrumental artifacts.

**Throughput correction**

We corrected the flux loss in PSF subtraction via negative disk injection, an essential step before extracting the disk spectrum. Negative injection is a common approach to calibrate the impact (i.e., flux loss) of PSF subtraction in high-contrast imaging[52]. In negative disk injection, a disk model at a given wavelength was created and convolved with a PSF model at the same wavelength. We used an empirical PSF measured from an unsaturated IFU observation ($R$~100) of TYC 4433-1800-1 from the commissioning program (PID 1128; PI: N. Luetzgendorf). We then subtracted the disk model from the science image at each wavelength, producing disk-free science images. After the negative injection, we performed RDI to subtract the stellar PSF in the science image. By minimizing the residuals of the disk-free science image, we determined the best-fit disk model at each wavelength with the Monte Carlo Markov Chain (MCMC) analysis using the emcee[53] package, under the framework of the DebrisDiskFM package[54].

We used a static geometric disk model to simulate the disk flux in each spectral channel for the throughput correction. In cylindrical coordinates, the density distribution of dust particles in a disk ring follows[55],

$$\rho(r,z) \propto \left[\left(\frac{r}{r_c}\right)^{-2\alpha_{\rm in}} + \left(\frac{r}{r_c}\right)^{-2\alpha_{\rm out}}\right]^{-\frac{1}{2}} e^{-\left(\frac{z}{hr^{\beta}}\right)^2},$$

where $r_c$ is the critical radius, $\alpha_{\rm in}$ is the rising component of the asymptotic power-law index, and $\alpha_{\rm out}$ is the decreasing component of the asymptotic power-law index. For the vertical structure of the disk, we adopted $\beta$ = 1 (linear flaring) and $h$ = 0.04[56]. The scattering phase function (SPF) of the disk is commonly approximated using the Henyey-Greenstein (HG) function[57] as

$$g_{\rm HG}(\theta) = \frac{1-g^2}{4\pi(1+g^2-2g\cos\theta)^{3/2}},$$

where $g$ is the HG asymmetry parameter and $\theta$ is the scattering angle. We used the anadisk_model[58] package, the Python-based disk modeling code, to create the disk model with four free parameters ($g$, $\alpha_{\rm in}$, $\alpha_{\rm out}$, and $f_{\rm model}$) per spectral channel. We included a flux scaling factor $f_{\rm model}$ because our disk modeling is only a ray-tracing approach.

We adopted other wavelength-independent disk parameters from ref.[24], which are the position angle of 99.1°, the disk inclination of 30.2°, the critical radius of 79.0 au, zero central offsets, and the distance of 47.72 pc[8]. We obtained the best-fit disk parameters at each wavelength by maximizing the log-likelihood function as follows:

$$\ln \mathcal{L}(\Theta|D_{\rm obs}) = -\frac{1}{2}\sum_{i=1}^{N_{\rm pix}}\left(\frac{D_{\rm obs,i}-D_{\rm model,i}}{\sigma_{\rm obs,i}}M_i\right)^2 - \sum_{i=1}^{N_{\rm pix}}\ln\sigma_{\rm obs,i}M_i - \frac{N_{\rm pix}}{2}\ln 2\pi,$$

where Θ is a set of disk parameters (that are, $g_{\rm HG}$, , $\alpha_{\rm in}$, $\alpha_{\rm out}$, and $f_{\rm model}$), $D_{\rm obs}$ is the disk image obtained with RDI, $D_{\rm model}$ is the disk model, and $\sigma_{\rm obs}$ is the uncertainty map at each wavelength. We masked the diffraction spikes and the horizontal negative stirps caused by saturation[44]. The uncertainty in each spatial pixel of $\sigma_{\rm obs}$ was estimated as the standard deviation in the spectral direction (100 channels) by using the disk-free residual cube. To increase the accuracy of the uncertainty estimation, the disk-free residual cube was obtained via an iteration process. In each iteration, we only updated the best-fit disk model.

We defined the throughput ($T_\lambda$) of our PSF subtraction at each wavelength as:

$$T_\lambda = \frac{(F_{\rm disk} - F_{\rm res})}{F_{\rm model}},$$

where $F_{\rm disk}$ is the disk flux measured in the disk image after RDI, $F_{\rm res}$ is the residual flux measured in the disk-free image after RDI, and $F_{\rm model}$ is the flux of the injected disk model. The numerator represents the flux change of the injected disk model caused by RDI PSF subtraction, assuming the two RDI reductions have the same throughput at the same position. The subtraction of $F_{\rm res}$ avoids the impacts of potential stellar speckles and potential residual disk flux that was not included in the disk model.

RDI provides the most robust recovery of disk morphology[52,59] in total intensity. By masking the disk region, we mitigated the oversubtraction effect[51]. As a result, for example, the estimated throughput is higher than 90% between 1–5 μm (Extended Data Fig. 2). The drop of throughput at <1.1 μm is probably caused by the spectral mismatch between the science and reference targets (Extended Data Fig. 1). Moreover, the throughput correction factors do not correlate with the shape of the disk reflectance spectrum shown in Fig. 1, indicating the robustness of the water-ice detection. Furthermore, we examined the possibility of a false positive of the water-ice gradient caused by the throughput correction. Similar to Fig. 1, we extracted the disk reflectance spectra at different distances but without applying the throughput correction and presented the spectra in Extended Data Fig. 4. The water-ice gradient is still visible with similar behaviors as in Fig. 1. Overall, the combination of multiple tests (i.e., Extended Data Figs. 1, 2, 4, and 5) demonstrate the robustness of post-processing by ruling out potential artifacts that can affect the results (i.e., the water ice detection and the radial gradient).

We used the ray-tracing approach[58,60] instead of a radiative transfer approach to create the disk model because we only focused on minimizing the residual of the disk-free image to determine the flux loss in RDI. The NIRSpec IFU observation is spatially under-sampled[44] and has limited spatial coverage in the azimuthal direction caused by eight diffraction spikes, which degrade the capability of NIRSpec IFU to characterize the disk properties in the spatial direction via radiative transfer approaches. It also limits the capability of the ray-tracing model to further constrain the disk properties. For example, the determination of the HG parameter is very sensitive to the coverage of the scattering angles, especially for the small scattering angles that are obstructed by spikes and instrumental artifacts (Fig. 1b). The lack of spatial information also prevents us from exploring the disk reflectance spectrum in the azimuthal direction. Therefore, we characterize the dust grain properties in the spectral direction by extracting the disk spectrum.

**Measurement of the disk reflectance spectrum**

To measure the disk reflectance spectrum, we performed aperture photometry in each spectral channel of the NIRSpec IFU after RDI. To avoid any potential angular dependence of scattering intensity[61] and contaminations from the residuals of eight bright diffraction spikes, we extracted the disk spectrum at the scattering angles around ~90° on both sides (east and west) of the disk. The strong diffraction spikes and instrumental artifacts also prevent us from performing detailed gradient analysis in the azimuthal direction. For example, regions at the scattering angles of around 60° (NE) and 115° (SW) are affected by the artifacts caused by detector saturation[44], shown as the dark horizontal strip in Fig. 1b. Regions at the scattering angles of around 60° (NW) and 115° (SE) are divided by additional diffraction spikes and the width of spikes becomes wider at longer wavelengths. Hence, there are not enough noncontaminated regions in the NW and SE directions to extract the disk spectrum. The apertures used to extract the disk spectrum are shown in Fig. 1b. The corresponding uncertainty of the disk spectrum was estimated as the standard deviation in the spectral direction (100 channels) using the disk-free residual cube and the same aperture.

Converting the observed disk flux reduced with RDI to the disk reflectance spectrum requires four flux corrections: (1) the throughput correction to account for RDI flux loss, (2) a correction term due to the stellar illumination, (3) a correction term to account for the PSF convolution, and (4) a correction term to remove the stellar color. The throughput correction was obtained via negative disk injection. We applied an $r^2$ scaling factor to correct the stellar illumination effect, where $r$ is the stellocentric distance. This illumination correction is a wavelength-independent correction. Because the debris disk around HD 181327 was spatially resolved and the instrumental PSF is wavelength-dependent, the PSF convolution effect needs to be corrected for using a fixed aperture in all wavelengths. A single disk model was created based on the best-fit disk model and convoluted with the unsaturated empirical PSF model (PID 1128; PI: N. Luetzgendorf). The correction factor of the PSF convolution effect is the ratio of the disk flux between the original model and the PSF convolved model at a given wavelength. The obtained correction factor follows a simple decreasing trend in the spectral direction, consistent with the trend of larger PSFs at longer wavelengths and uncorrelated with the observed water-ice features. The difference in the correction factor between the shortest and longest wavelengths is less than 10%.

Due to saturation issues, we did not observe the stellar spectrum of HD 181327 with the NIRSpec IFU. To remove the stellar color and thus reveal the scattering efficiency of the dust, we used the BT-Settl[62] model spectrum of HD 181327 with stellar parameters of $T_{\mathrm{eff}}$ = 6,400 K, $\log(g)$ = 4.5 cm s$^{-2}$, and zero metallicity. For comparison, HD 181327 has an effective temperature of $6{,}375^{+2.7}_{-2.1}$ K with the stellar surface gravity of $4.23^{+0.01}_{-0.01}$ cm s$^{-2}$ and the metallicity of $-0.23^{+0.08}_{-0.04}$, adopted from ref.[63]. The stellar photosphere model was smoothed to match the spectral resolution ($R\sim100$) of the IFU observation. By dividing the extracted disk spectrum by the stellar spectrum in the spectral direction, we can remove the stellar color.

After applying four flux corrections, we obtained the disk reflectance spectrum extracted from a given stellocentric distance at 0.6–5.3 μm with a spectral resolution of ~100. Finally, we trimmed the first 20 (<0.7 μm) and last 20 (>5.2 μm) channels and normalized the reflectance spectrum with respect to its spectral reflectance at 2.5 μm, enabling the comparison.

We present the comparison of the disk spectrum at 90–105 au and the stellar photosphere model in Extended Data Fig. 3, showing the detection of the water-ice feature at ~3 μm before and after the correction of stellar color. The overall slope of the disk spectrum is slightly shallower than the stellar photosphere model with clear changes in slopes between 2.7 and 3.4 μm, suggesting a change in the scattering efficiency of dust particles in the disk due to water ice, as presented in the reflectance spectrum.

**Gradient measurement**

To investigate the gradient in the radial direction, we used three spectral extracting regions (Fig. 1) to probe the spectral gradients in the middle ring (region 1; 80–90 au), outer ring (region 2; 90–105 au), and halo (region 3; 105–120 au), respectively. The inner edge of region 1 is about 80 au which is close to the inner edge (76 au) of the disk ring[24,33]. We adjusted the size of the extracting regions to avoid the noisy regions (e.g., <80 au) after PSF subtraction and residuals from diffraction spikes. Due to the exponential decrease of the disk surface brightness, we only added one region to probe the disk halo between 105–120 au. We performed the dust model fitting for disk reflectance spectra extracted from three regions. The fitted models are shown in Fig. 1 and the derived dust grain properties are listed in Table 1. We also extracted the disk spectra from the two sides (east and west) of the disk in each spectral extracting region, shown in Extended Data Fig. 5. The reflectance spectra have no significant differences between the two sides of the disk.

**Dust grain model**

The model reflectance at a given wavelength was calculated as follows:

$$\frac{F_{\rm disk}}{F_{\rm star}} = f \sum_m f_{\rm pop} \int_{a_{\rm min}}^{a_{\rm max}} g_{\rm HG}(\theta) Q_{\lambda,{\rm sca},m} \pi a^{2+p}\, da,$$

where $F_{\rm star}$ is the stellar flux, $f$ is the scaling factor, $f_{\rm pop}$ is the dust population fraction, $a$ is the grain radius, $p$ is the power-law size distribution, and $Q_{\rm sca}$ is the scattering efficiency. We adopted the scattering phase function of $g_{\rm HG}(90°)$ with $g$ = 0.3[10]. We used Mie theory[64] to calculate the scattering efficiency of a given medium using the Miepython[65] package. The number of dust populations is denoted by $m$. In the case of two dust populations, the fraction of the first and the second populations are represented by $f_{\rm pop1}$ and $1 - f_{\rm pop1}$, respectively. If there is only one dust population, the fraction is 1. In each dust population, several dust species (i.e., $H_2O$, FeS, olivine, etc.) were mixed to form a homogeneous effective medium. The optical constant ($n_{\rm eff}$) of the effective medium can be calculated using the Bruggeman rule[66], which is written as:

$$\sum_{s=1}^{S} V_s \frac{n_s^2 - n_{\rm eff}^2}{n_s^2 + 2n_{\rm eff}^2} = 0 \text{ and } \sum_{s=1}^{S} V_s = 1,$$

where $s$ is the dust component index, $S$ is the total number of components, $V_s$ is the volume fraction of a given component, and $n_s$ is the optical constant of a given component. The porosity of the medium is also included as one of the dust components with the optical constant of vacuum. We adopted the lab measurements of optical constants for amorphous and crystalline water ice[67], FeS[68], olivine ($MgFeSiO_4$, i.e., $Fo_{100}$)[69], amorphous carbon[70], pyroxene ($MgSiO_3$, $Mg_{0.7}Fe_{0.3}SiO_3$, and $Mg_{0.4}Fe_{0.6}SiO_3$)[69], and SiC[71]. The free parameters are $f$, $a_{\rm min}$, $a_{\rm max}$, $p$, and $V_s$. The optical constants of water-ice are measured at 50 K[67], consistent with the dust temperature at 90–120 au in Extended Data Fig. 6.

Mie theory is the simplest approximation that assumes a spherical shape for dust grains. Mie theory was commonly used to set up expectations in debris disk modeling. We recognized that Mie theory is probably inadequate because it is inadequate for comets. However, in debris disks, especially for HD 181327, we do not have definitive evidence about its grain shapes. As a result, we stay with the simplest Mie theory in this work.

We fitted the measured disk reflectance spectrum with models of reflectance spectrum using scipy.optimize.curve_fit[72] and the emcee[53] package. The short wavelength limit of 1.1 μm is determined by the shortest wavelength in the laboratory measurement of optical constants of water ice[67]. We first used scipy.optimize.curve_fit to obtain a good fit and used it as initial input in the MCMC analysis. We then performed the MCMC analysis by maximizing the log-likelihood function to obtain the posterior distributions of the dust parameters, listed in Table 1 and Extended Data Table 1. The corner plots of the dust parameters are presented in Extended Data Figs. 9–11, showing the distribution of free parameters in models presented in Table 1. The reported 68% uncertainty corresponds to the commonly used 1 $\sigma$ confidence level, which is consistent with the confidence level adopted throughout this work.

To keep the model simple and minimize overfitting, we started with the model using three dust compositions to fit the disk reflectance spectrum, including two phases (amorphous and crystalline) of water ice and one refractory dust species (FeS or olivine). The presence of FeS and olivine in the HD 181327 disk was previously inferred by refs. [11,20,24]. We added the other dust species one at a time and only kept the dust species that contributed to the model fitting. We found that using a mixture of FeS and olivine (in the first population) can lower the reduced $\chi^2$ by 0.1–0.2 compared to using only one of the two species. We used the same dust species across three disk regions. We found four dust species (water ice, FeS, olivine, and pyroxene) can reproduce the observed disk reflectance spectra. Due to the similarity of the olivine and pyroxene spectra at 1–5 μm, we used only olivine in the dust model to keep the simplicity of the model because our data cannot distinguish them.

We tried the single dust population first and found that we obtained a good fit for the reflectance spectrum at 80–90 au but not for the spectra at 90–105 au and 105–120 au. For example, we find that the dust model (model #4; 90–105 au) using one dust population cannot fully reproduce the water-ice feature at 3 μm, leaving significant deviations at around 2.8 μm and minor deviations at around 3.3 μm (Extended Data Fig. 3). Including additional dust species (i.e., amorphous carbon or SiC) did not improve the spectral fitting. Another way of improving the spectral fitting is adopting a more complicated model with multiple dust populations. Hence, we proposed using two dust populations to fit the spectral reflectance, with one accounting for the water-ice-depleted population and another accounting for the water-ice-rich population. In fact, it is expected to have both water-ice-depleted and water-ice-rich dust at >90 au because the radiation pressure from the host star can blow out small and ice-free dust grains from the inner part (<90 au) to the outer part of the disk (>90 au). After including two dust populations in model #2 and shown in Extended Data Fig. 3, the water-ice feature at 3 μm can be properly fitted. As a result, the $\chi_{\nu}^2$ is significantly reduced from 2.8 to 1.6 (see panel c in Extended Data Fig. 3). Furthermore, the best-fit volume fractions of $H_2O$ in the first dust population are close to zero (Table 1), consistent with the use of a second water-ice-rich dust population.

The difference between the single and two dust populations is whether we mixed water ice with the rest of the materials using the Bruggeman rule. In other words, the model of two dust populations considers two separated dust mediums, and the model of a single dust population assumes a well-mixed dust medium. This difference gives the different spectral features of water ice at around 3 μm shown in Extended Data Fig. 3.

The obtained dust model also predicts the mass fraction of water ice in the disk. The mass fraction of water ice at 80–90 au can be directly obtained based on the best-fit volume fractions, which is about 0.001. As for models at 90–105 au and 105–120 au, the mass fraction of water ice can be derived by using the integration of the corresponding size distributions[73] and the population fraction, yielding the water-ice mass fractions of about 0.075 and 0.210 at 90–105 au and 105–120 au, respectively. In the calculation, we adopted the densities of $H_2O$ ice, FeS, and olivine to be 0.92, 4.83, and 3.71 g cm$^{-3}$, respectively. Previous SED fitting results gave the water-ice mass fractions of 23.8%[11] or 40.8%[20]. Although the extracted disk regions and observing wavelengths are different in our work from previous SED fittings, the derived mass fraction is similar.

The limited spectral coverage (i.e., 1–5 μm), low spectral resolutions (i.e., R~100), and the uncertainty of NIRSpec data as well as the lack of spectral features of minerals in our data result in a certain level of model degeneracy, preventing us from unambiguously identifying dust compositions via their unique spectral features, except water ice. A similar degeneracy problem was also presented in the analysis of the reflectance spectra of comet 67P in the range 0.3–5.1 μm[74]. To understand the model degeneracy in dust composition, we made a degeneracy test by performing the fitting on fake model spectra created with known dust parameters. We created model spectra similar to model #1 or #2 but with different volume fractions and adopted the same uncertainty from the observed spectra (Fig. 1). The test result shows that the recovered dust properties are consistent with the synthetic model if we use the same dust species. However, our test failed to recover the consistent dust properties if two different dust species have very similar spectra or if a dust species is flat and featureless. For example, olivine and pyroxene spectra at 1–5 μm are very similar. We can replace olivine with pyroxene to produce a very similar model. Amorphous carbon is one of the most common dust species and has a flat and featureless spectrum at 1–5 μm. Our degeneracy test finds that we cannot accurately determine the presence and abundance of amorphous carbon. We note that the main results of this work (water-ice detection, water-ice gradient, and spectral analogy (removing the disk continuum using the ice-free spectrum <90 au instead of a continuum model)) are independent of the modeling. To further constrain the dust composition, combined modeling of all scattering properties (reflectance, phase function, and degree of polarization) together with the SED may help to break some of those degeneracies. Nevertheless, observing unique solid-state features is necessary to unambiguously identify the composition of both icy and refractory grains.

**Thermal sublimation timescale of water ice**

To calculate the mass loss rate of water ice ($\dot{m}$, in g cm$^{-2}$ s$^{-1}$) due to thermal sublimation, based on radiative equilibrium, we simultaneously solved for the temperature of the dust grain ($T_{\rm gr}$, in K) and included the effect of latent heat release following[75]

$$\left(\frac{R_*}{D}\right)^2 \int_0^\infty Q_{{\rm abs},\lambda,a} \pi B_\lambda(T_*) d\lambda = 4\left[\int_0^\infty Q_{{\rm abs},\lambda,a} \pi B_\lambda(T_{\rm gr}) d\lambda + \eta H \dot{m}(T_{\rm gr})\right],$$

where $R_*$ is the stellar radius, $T_*$ is the stellar temperature, $D$ is the grain distance, $H = 2.45 \times 10^{10}$ erg g$^{-1}$ is the specific heat of sublimation[76], $B_\lambda$ is the Planck function, and $Q_{\rm abs}$ is the absorption efficiency. We calculated $Q_{\rm abs}$ using the Miepython package[65] under the assumption of Mie scattering. We used the optical constant of crystalline water ice from ref.[77] because the previous one[67] used in grain modeling has a lower wavelength limit of 1.1 μm. We calculated the absorption efficiency for a dirty-ice model by mixing water ice ($V_{\rm H2O} = 0.23$) with olivine ($V_{\rm olivine} = 0.03$) and porosity ($V_{\rm porosity} = 0.74$).

The mass loss rate of water ice per unit of surface area ($\dot{m}$, in g cm$^{-2}$ s$^{-1}$) is given by

$$\dot{m}(T_{\rm gr}) = P_{\rm s}(T_{\rm gr}) \sqrt{\frac{\mu}{2\pi k T_{\rm gr}}},$$

where $P_{\rm s}$ is the saturation vapor pressure, $\mu = 2.99 \times 10^{-23}$ g is the molecular mass of the water, and $k$ is the Boltzmann constant. We adopted the relation of $P_{\rm s}$ from the lab measurement of water ice[78], which is given by

$$P_{\rm s}(T_{\rm gr}) = 10^{-2403.4/T_{\rm gr} + 9.183837},$$

where the unit of $P_{\rm s}$ is torr.

The erosion rate of an icy grain caused by thermal sublimation ($\dot{s}_{\rm sub}$, in cm s$^{-1}$) is

$$\dot{s}_{\rm sub} = \frac{\eta \dot{m}(T_{\rm gr})}{\rho},$$

where $\eta$ is the fraction of the surface covered by sublimating materials and $\rho$ is the density of the grain material. Hence, the thermal sublimation timescale ($t_{\rm sub}$, in s) of water ice is

$$t_{\rm sub} = \frac{a}{\dot{s}_{\rm sub}} = \frac{a\rho}{\eta \dot{m}(T_{\rm gr})},$$

where $a$ is the grain radius. As a one-order-of-magnitude approximation, we adopted $\eta$ to be 1, which may be an overestimation.

We calculated and presented the dust temperature of a dirty ice model and corresponding thermal-sublimation timescales for different grain radii in Extended data Fig. 6. The dust temperature is around 50 K at 80–120 au for grain sizes of 1–10 μm. The sublimation timescale indicates that thermal sublimation is inefficient in destroying water ice in the HD 181327 disk at >80 au. As for pure water ice that is less absorbing, the absorption efficiency is much lower at wavelengths of 0.2–1 μm, resulting in a lower dust grain temperature and a longer sublimation timescale. Therefore, we exclude thermal sublimation as the mechanism of destroying water ice at >80 au.

**Photodesorption timescale of water ice**

Ultraviolet (UV) photons absorbed by an icy grain can not only dissociate the water molecules but also lead to molecule desorption from the grain surface, which is so-called photodesorption or photosputtering[75]. The outer part of the debris disk may be exposed to the UV radiation from the host star if there is no inner disk shielding the light. In the case of HD 181327, the ALMA observation[12] and the NACO observation using sparse aperture masking[79] did not find an inner disk, indicating the important role of UV photodesorption in removing water ice from the disk.

The erosion rate of an icy grain caused by photodesorption ($\dot{s}_{\rm uv}$, in cm s$^{-1}$) is calculated as[75]

$$\dot{s}_{\rm uv} = \frac{\eta m_{\rm H_2O} Y N_{\rm abs}}{4\rho},$$

where $m_{\rm H_2O}$ = 2.99 × 10$^{-23}$ g is the mass of a water molecule, $Y$ is the desorption probability, and $N_{\rm abs}$ is the number of absorbed photons. We calculated the number of absorbed photons between the UV wavelength range of 0.091 μm (13.6 eV) and 0.24 μm (5.1 eV) as

$$N_{\rm abs} = \int_{\lambda_{\rm min}}^{\lambda_{\rm max}} \frac{F(D,\lambda)}{hc/\lambda} Q_{{\rm abs},\lambda} d\lambda,$$

where $F(D,\lambda)$ is the stellar flux at the location of the grain, $h$ is the Planck constant, $c$ is the speed of light, and $Q_{\rm abs}$ is the absorption efficiency of a given medium. To explore the absorption efficiency at the UV wavelengths, we used the optical constant of crystalline water ice from ref. [77] because the previous one[67] used in grain modeling has a lower wavelength limit of 1.1 μm. We calculated the absorption efficiency for a dirty-ice model by mixing water ice with olivine and porosity (model #2 in Table 1). Unlike $Q_{\rm abs}$ for pure water ice that becomes «1 between 5.1–7 eV, $Q_{\rm abs}$ for dirty ice is close to 1 in the entire UV range, which is expected as the mixture becomes much more absorbing (i.e., higher imaginary part in $n_{\rm eff}$). The integral of $Q_{\rm abs}$ is not very sensitive to the grain size (i.e., 1–100 μm) in our UV range. Therefore, we adopted a grain size of 10 μm for illustration in our one-order-of-magnitude approximation.

Finally, the photodesorption timescale ($t_{\rm uv}$, in s) of water ice is

$$t_{\rm uv} = \frac{a}{\dot{s}_{\rm uv}} = \frac{4\rho a}{\eta m_{\rm H_2O} Y N_{\rm abs}}.$$

We adopted the desorption probability of crystalline water ice[80,81] to be 2 × 10$^{-3}$ and $\eta$ to be 1. Assuming a constant erosion rate, the photodesorption timescale is linearly proportional to the grain size, with larger grains taking more time to erode.

Based on the TD1 catalog[82], the apparent fluxes of HD 181327 are 5.6 × 10$^{-5}$, 5.2 × 10$^{-4}$, 1.52 × 10$^{-4}$ erg cm$^{-3}$ s$^{-1}$ at 156.5, 196.5, and 236.5 nm, respectively. The Far Ultraviolet Spectroscopic Explorer (FUSE) observed HD 181327 (Prog. ID: F320; PI: Brown), providing a spectrum at 0.09–0.12 μm with a flux level of around 2 × 10$^{-6}$ erg cm$^{-3}$ s$^{-1}$ or less. Such a flux level is much lower than the UV flux at longer wavelengths. Due to the potential time variability of stellar UV flux, as a first-order approximation, we performed a simple interpolation using interp1d in the SciPy package[72] based on the UV measurements between 0.09 and 0.24 μm. The obtained number of absorbed photons ($N_{\rm abs}$) is dominated by the UV flux at longer wavelengths (i.e., 0.15–0.24 μm).

We then calculated and presented the photodesorption timescale in Extended Data Fig. 7, assuming the optical depth of 0. Even for a large particle of 100 μm, the photodesorption timescale is still much less than the system age and close to the collision timescale, indicating its efficiency in removing water ice from dust grains. However, we assumed the fraction ($\eta$) of the grain surface covered by water ice equals one, which is likely an overestimation, leading to an underestimation of the photodesorption timescale.

To explain the observed water-ice gradient, if water ice is removed by photodesorption, we assume the optical depth ($\tau$) is nonnegligible in the radial direction of the disk and follows the profile of the disk optical depth (i.e., Fig. 4 in ref. [24]). In other words, we assume the disk is not optical thin in the radial direction. Because the inner edge of the HD 181327 disk was affected by the residual of PSF subtraction, we adopted the exponent $\alpha_{\rm in}$ in the density distribution of disk to be 16.2 from ref. [24]. We adopted the value of exponent $\alpha_{\rm out}$ to be –7 from our best-fit models.

The incident stellar flux ($F$) after the UV extinction can be calculated based on the stellar flux at the grain distance without the extinction ($F_0$) as

$$F = F_0 e^{-\tau}.$$

The total optical depth ($\tau$) is calculated in a range of 70–120 au, which is given by

$$\tau = \int_{70}^{120} \tau_D \, dD,$$

with

$$\tau_D = (D/r_{\rm c})^{\alpha_{\rm in}} \ \text{if} \ D \leq r_{\rm c},$$

or

$$\tau_D = (D/r_{\rm c})^{\alpha_{\rm out}} \ \text{if} \ D > r_{\rm c},$$

where $\tau_D$ is the optical depth at a given grain distance ($D$) and $r_{\rm c}$ is the critical radius of 79.0 au. The inner edge of the HD 181327 disk is approximately 70 au[24]. By assuming the total UV extinction in a range of 70–120 au, we can calculate the optical depth at a given grain distance, thus obtaining the incident stellar flux after the UV extinction at a given distance. After considering the potential UV extinction, we calculated and presented the photodesorption timescales for a 10 μm grain with different optical depths in Fig. 2.

**Collison timescale**

Fig. 2 shows the estimated collision timescale ($t_{\rm coll}$, in yr) of dust grains, which is given by[83,84]

$$t_{\rm coll} = 3000 \ {\rm yr} \left(\frac{D}{30 \ {\rm au}}\right)^{7/2} \left(\frac{M_{\rm submm}}{0.1 \ M_{\oplus}}\right)^{-1} \left(\frac{\sqrt{a_{\rm min} a_{\rm max}}}{1 \ {\rm mm}}\right) \left(\frac{M_{\odot}}{M_*}\right)^{1/2},$$

where $D$ is the stellocentric distance, $M_{\rm submm}$ is the dust mass derived from submillimeter observations, $a_{\rm min}$ is the minimum grain size, $a_{\rm max}$ is the maximum grain size, and $M_*$ is the mass of the host star. We adopted the dust mass, minimum and maximum grain sizes, and the stellar mass to be 0.422 $M_{\oplus}$[12], 1.3 μm, 1 cm[12], and 1.3 $M_{\odot}$[85], respectively. At 90 au, $t_{\rm coll}$ is about 3,325 yr. However, it is important to note that estimating the collision timescale is a complex issue and this calculation is just a one-order-of-magnitude approximation.

As an important source of newly generated icy grains, the collision timescale sets the highest replenishment rate of water ice if we assume fragments produced by the collision of the icy parent bodies always contain water ice on the surface. It is well established that the dust grains in the debris disk are not primordial, but that they should be continuously replenished by the collision of bigger objects[1]. Therefore, the detection of water ice in the disk implies that the stellar age sets the lower limit of the replenishment rate of water ice. By comparing the erosion rate of water ice (due to sublimation and/or photodesorption) with its replenishment rate, we can make a qualitative expectation about whether we can detect water ice, thus qualitatively explaining the observed water-ice gradient.

**Dust grain blowout size**

To evaluate the minimum grain size obtained with dust modeling, we calculated the dust grain blowout size ($a_{\text{blow}}$) assuming a spherical shape for the particles. The blowout size is determined when the force of radiation pressure equals the gravitational attraction, which is given by[86,87]

$$a_{\text{blow}} = \frac{6L_* \langle Q_{\text{pr}} \rangle}{16\pi\rho G M_* c}.$$

Here, $L_*$ is the host star luminosity, $\langle Q_{\text{pr}} \rangle$ is the average radiation-pressure efficiency, $\rho$ is the density of a dust grain, $G$ is the gravitational constant, and $c$ is the speed of light.

Adopting the best-fit dust parameters from Table 1, we calculated the average radiation-pressure efficiency using

$$\langle Q_{\text{pr}} \rangle = \frac{\int_0^\infty Q_{\text{pr},\lambda,a} F_*(\lambda) d\lambda}{\int_0^\infty F_*(\lambda) d\lambda},$$

where $F_*(\lambda)$ is the stellar flux adopted from the stellar photosphere model and $Q_{\text{pr},\lambda,a}$ is the radiation-pressure efficiency of a given grain mixture and size at a given wavelength. We calculated $Q_{\text{pr},\lambda,a}$ using the Miepython package[65]. We adopted the densities of $H_2O$ ice, FeS, and olivine to be 0.92, 4.83, and 3.71 g cm$^{-3}$, respectively. We calculated the density of a grain mixture based on the best-fit volume fractions of dust species and porosity.

For the first dust population, the obtained grain blowout sizes are 0.95, 0.96, and 1.67 μm for models #1, #2, and #3 in Table 1, respectively. For the second dust population, the obtained grain blowout sizes are 0.21 and 1.43 μm for models #2 and #3, respectively. The minimum grain sizes in models #1 and #2 are larger than the blowout size as expected. However, the minimum grain size in model #3 is smaller than the corresponding blowout size. The possible presence of sub-blowout particles in the halo may lead to a significant contribution in the disk spectrum at 105–120 au, resulting in $a_{\text{min}} < a_{\text{blow}}$[88].

**Comparison of the water-ice feature**

The New Horizons spacecraft provided the first detection and measurements of our Solar System's outer Kuiper Belt debris disk[89,90], revealing that over 95% of the dust mass orbiting our Sun is concentrated in this disk. The dust in this disk originates from collisions among Kuiper Belt objects and sputtering of interstellar medium dust[90,91]. The surface particles of minor bodies are embedded in an extremely porous matrix, analogous to the vacuum of interplanetary space rather than uniform solid slabs[92–94].

The HD 181327 disk is composed of dust particles generated by collisions among planetesimals beyond the snowline[1]. The composition of the dust is expected to reflect that of the parent planetesimals if the dust remains unaltered as it was released. However, the disk spectral continuum could be different from that of individual KBOs because individual KBOs may have different dust compositions. Furthermore, the dust ejection process could cause the grain size and size distribution to be different than those on the individual KBO surface, leading to a difference in their spectral continuum. Therefore, subtracting the continuum and only comparing the spectral features of a given dust species in the icy disk to that of known icy bodies in the Solar System is reasonable to see if there is any correlation and similarity. It could help us better understand the origin of the dust in the disk beyond the snowline.

The spectral shape of the water ice feature is affected by the underlying continuum. The disk reflectance spectrum shows the scattering efficiency of dust particles. A mixture of water-ice particles with other dust species (i.e., FeS, olivine, etc.) will lead to a change in the scattering efficiency of ice grains, thus changing the spectrum continuum. For example, the overall slope of the disk spectra is decreased as a function of distances, which is caused by the decrease in the minimum grain size. Furthermore, the flux jump in the spectral continuum (i.e., >90 au) between <2.7 and >3.4 μm can be reproduced with a mixture of olivine of FeS simply because their scattering efficiency has a similar jump around 3 μm (i.e., Extended Data Figure 8).

To better inspect and compare the water ice feature, we subtracted the disk continuum contributed by FeS and olivine (i.e., the continuum slope and the continuum flux jump around 3 μm; see Extended Data Figure 8). We first created a continuum model of a FeS and olivine mixture by replacing water ice with vacuum, thus increasing the porosity of the dust grain, following the best-fit dust parameters (models #2 and #3) from Table 1. Then we subtracted the continuum dust model from the measured disk reflectance spectrum, obtaining the water-ice-dominated spectrum. Among spectra of minor bodies in our Solar System[15,16,34,35], we found some of the water-ice-rich KBOs[17,36] have the closest spectra of the water ice feature at 3 μm to the spectra of the HD 181327 disk (i.e., Extended Data Fig. 8). We present the spectral comparison in Fig. 3.

Removing the continuum is a tool for better looking at spectral features in the residuals. A similar approach of removing the known dominated components in the spectrum for better characterizing the remaining faint spectral features has been used in analyzing the reflectance spectrum of protoplanetary disks (e.g., ref. [95]). By only removing the same amount of FeS and olivine suggested by the model fitting, we properly preserved the water-ice feature and avoided potential oversubtraction.

We did not perform the same continuum removal for normalized icy body spectra because the normalized icy body spectra (2.5–5 μm) presented in Fig. 3 are dominated by icy features (i.e., icy KBO spectra fitting in Fig. 2 of ref. [15]). This is because silicates are featureless and relatively flat at 2.5–5 μm. As shown in Fig. 3, the normalized icy body spectrum is significantly affected by a change in the continuum slope only at around <2 μm. Therefore, the continuum of icy bodies contributed by minerals (i.e., silicates) does not significantly affect the water ice feature at 3 μm, thus not affecting the spectral comparison.

We scaled the distance of the disk based on the stellar UV luminosity. We chose the UV luminosity instead of the stellar temperature because HD 181327 is young and both disk and icy bodies in the Solar System are beyond the snowline. To calculate the solar UV luminosity, we adopted the median solar UV spectra from the SOlar Stellar Irradiance Comparison Experiment instrument[96], ranging from 115–240 nm. We obtained the UV luminosity of HD 181327 in the same wavelength range and performed the UV extinction correction by adopting the extinction model from Fig. 2. If we assumed an optical depth of 3 in the disk radial direction between 70–120 au, the total UV extinction from 70–90 au reduced the UV flux down to 13.3%, resulting a UV luminosity ratio of HD 181327 and the Sun to be about 23.1. Hence, the disk distance was scaled by a factor of 4.8 (Fig. 3). Different optical depths will lead to different scaling factors, but the difference is less than a factor of 2 if the optical depth is in a range of 1–5.

## Data availability

The original data used in this work is part of the GO programme 1563 (PI: C. Chen) and is publicly available from the STScI MAST archive (https://mast.stsci.edu). The specific observations analyzed can be accessed via doi:10.17909/mr4p-v151. Reduced data cube and extracted disk spectra used in the analysis are publicly available at Zenodo https://doi.org/10.5281/zenodo.14985028.

## Code availability

The post-processing and dust reflectance modeling codes used in this work are developed by C.X., as detailed in Methods. The post-processing process uses the jwstIFURDI package, which can be found at https://github.com/ChenXie-astro/jwstIFURDI.git. The dust reflectance modeling process uses the DDRM package, which can be found at https://github.com/ChenXie-astro/DDRM.git. Solving the complex numbers in the Bruggeman rule uses the mpmath package (https://mpmath.org/). The Mie scattering calculation uses the miepython package, which can be found at https://github.com/scottprahl/miepython.git. The disk 2D model uses the publicly available anadisk_model package, which can be found at https://github.com/maxwellmb/anadisk_model.git. The fitting procedure for the disk model uses the publicly available Markov chain Monte Carlo Ensemble sampler, the emcee package, which can be found at https://github.com/dfm/emcee.git, and the disk forward modeling code, the DebrisDiskFM package (https://github.com/seawander/DebrisDiskFM.git). In addition to emcee, Scipy (https://scipy.org/) is also used in the fitting procedure for the dust reflectance spectrum. Figures were made with Matplotlib v.3.3.0. under the Matplotlib license at https://matplotlib.org/.

**Acknowledgments** Nature thanks the anonymous reviewers for their contribution to the peer review of this work. This work is based on observations made with the NASA/ESA/CSA James Webb Space Telescope. The data were obtained from the Mikulski Archive for Space Telescopes at the Space Telescope Science Institute, which is operated by the Association of Universities for Research in Astronomy, Inc., under NASA contract NAS 5-03127 for JWST. These observations are associated with program #01563. Support for program #01563 was provided by NASA through a grant from the Space Telescope Science Institute, which is operated by the Association of Universities for Research in Astronomy, Inc., under NASA contract NAS 5-03127. B.T.B. is supported by an appointment to the NASA Postdoctoral Program at the NASA Goddard Space Flight Center, administered by Oak Ridge Associated Universities under contract with NASA. S.K.B is supported in part by an STScI Postdoctoral Fellowship. C.M. Lisse would like to acknowledge his support for this work from the Johns Hopkins University Applied Physics Laboratory sabbatical program and the NASA New Horizons mission project. N.P.- A. acknowledges funding through the ATRAE program of the Ministry of Science, Innovation, and Universities (MICIU) and the State Agency for Research (AEI) in Spain.

**Author contributions** C.X. led the program and performed the data reduction, PSF subtraction, spectrum modeling, and data analysis. C.X. wrote the manuscript. C.C. co-led the program and assisted with writing and spectrum modeling. T.B., C.I. and K.W. assisted with the preprocessing of the data. S.K.B., M.D.P., L.P. and T.B. assisted with the post-processing of the data. T.B. provided the cleaned empirical PSF model. S.G.W. coordinated the result comparisons with NIRCam observations. C.C., C.M.L., D.C.H., N.PA., A.G., B.T.B., J.A.S. and S.G.W. contributed to the interpretation of results. N.PA. provided the spectra from the DiSCo program. C.C., T.B., A.G., J.M.L., C.M.L., M.D.P., L.P., J.A.S. and S.G.W. contributed to obtaining the JWST data. All authors participated in the discussion of the results and/or commented on the manuscript.

**Competing interests** The authors declare no competing interests.

**Additional information**
**Correspondence and requests for materials** should be addressed to Chen Xie.

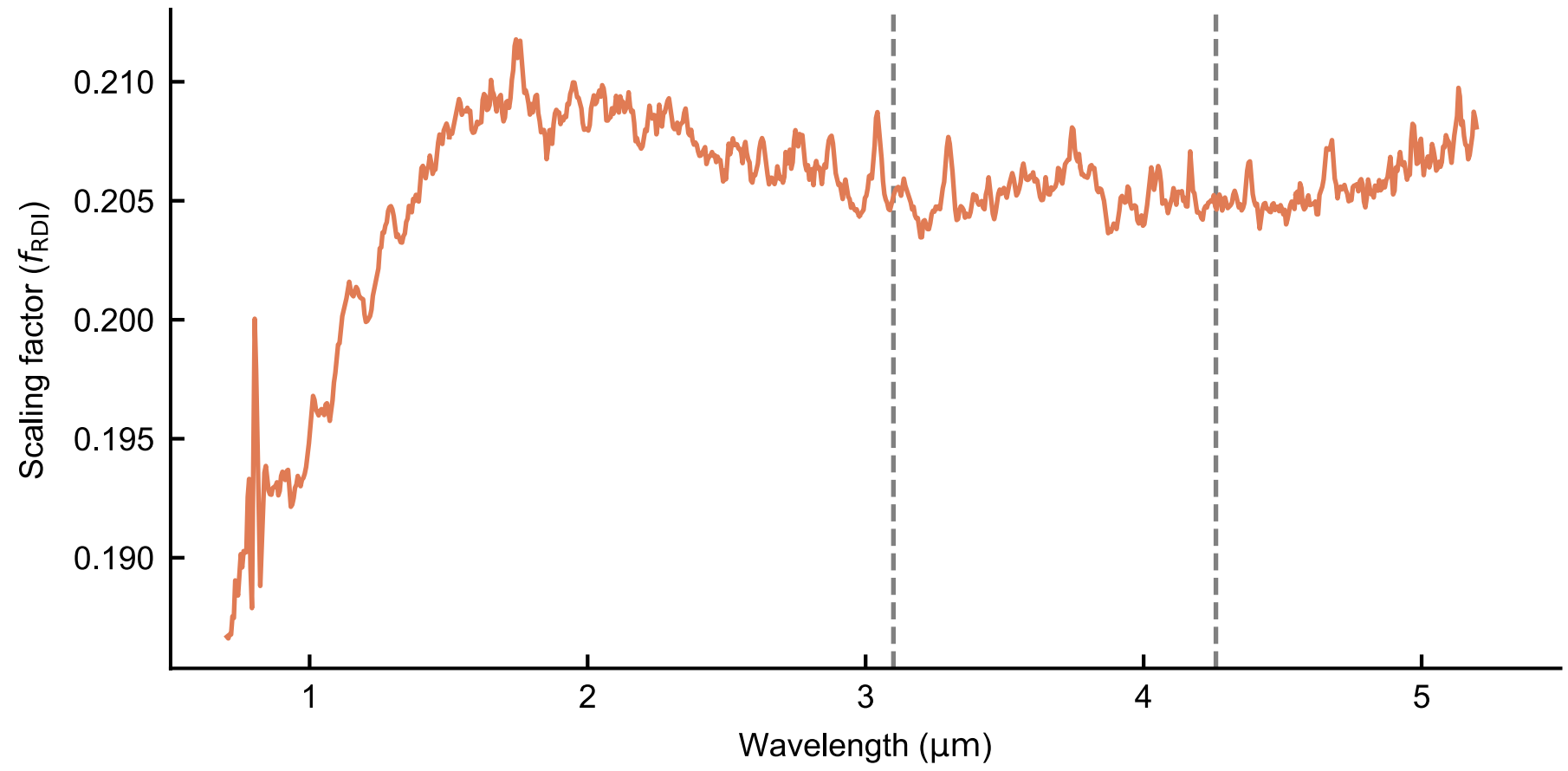

**Extended Data Fig. 1 | Scaling factor in RDI as a function of wavelength.** The scaling factor used in RDI is stable between 1.5–5 µm. The 2% drop below 1.5 µm is probably caused by the mismatch in spectral types of HD 181327 (F6[7]) and Iota Mic (F2[97]). Vertical dashed lines mark the Fresnel peak of water ice at 3.1 µm and the $CO_2$-ice feature at 4.25 µm.

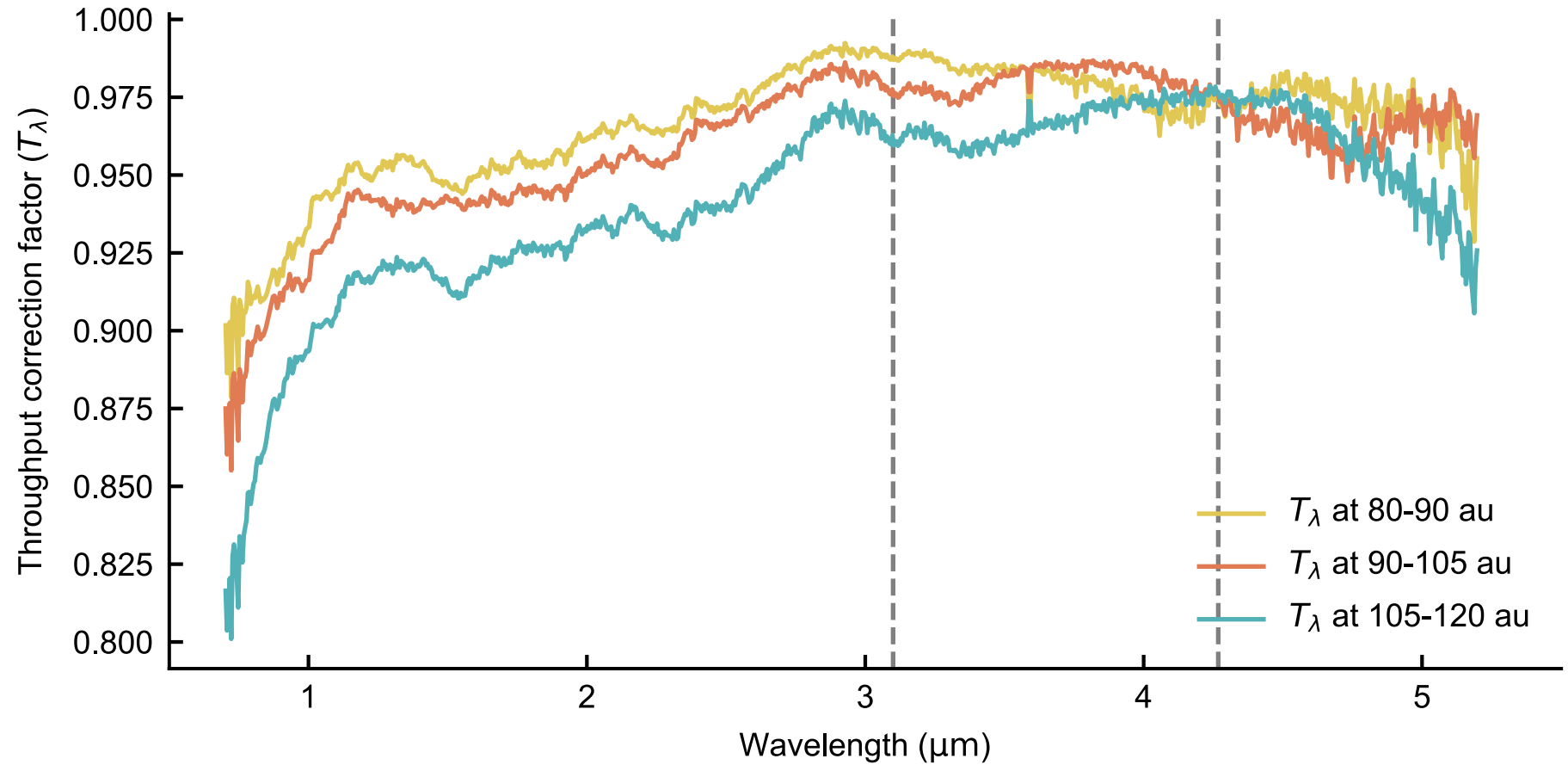

**Extended Data Fig. 2 | Throughput correction factors of RDI PSF subtraction.** The RDI throughputs of different disk-extracting regions as a function of wavelength. The flux loss caused by PSF subtraction is less than 10% and relatively stable between 1–5 µm. The drop in throughput at <1.1 µm is probably caused by the mismatch of spectral types between the science and reference stars. Vertical dashed lines mark the Fresnel peak of water ice at 3.1 µm and the $CO_2$-ice feature at 4.25 µm.

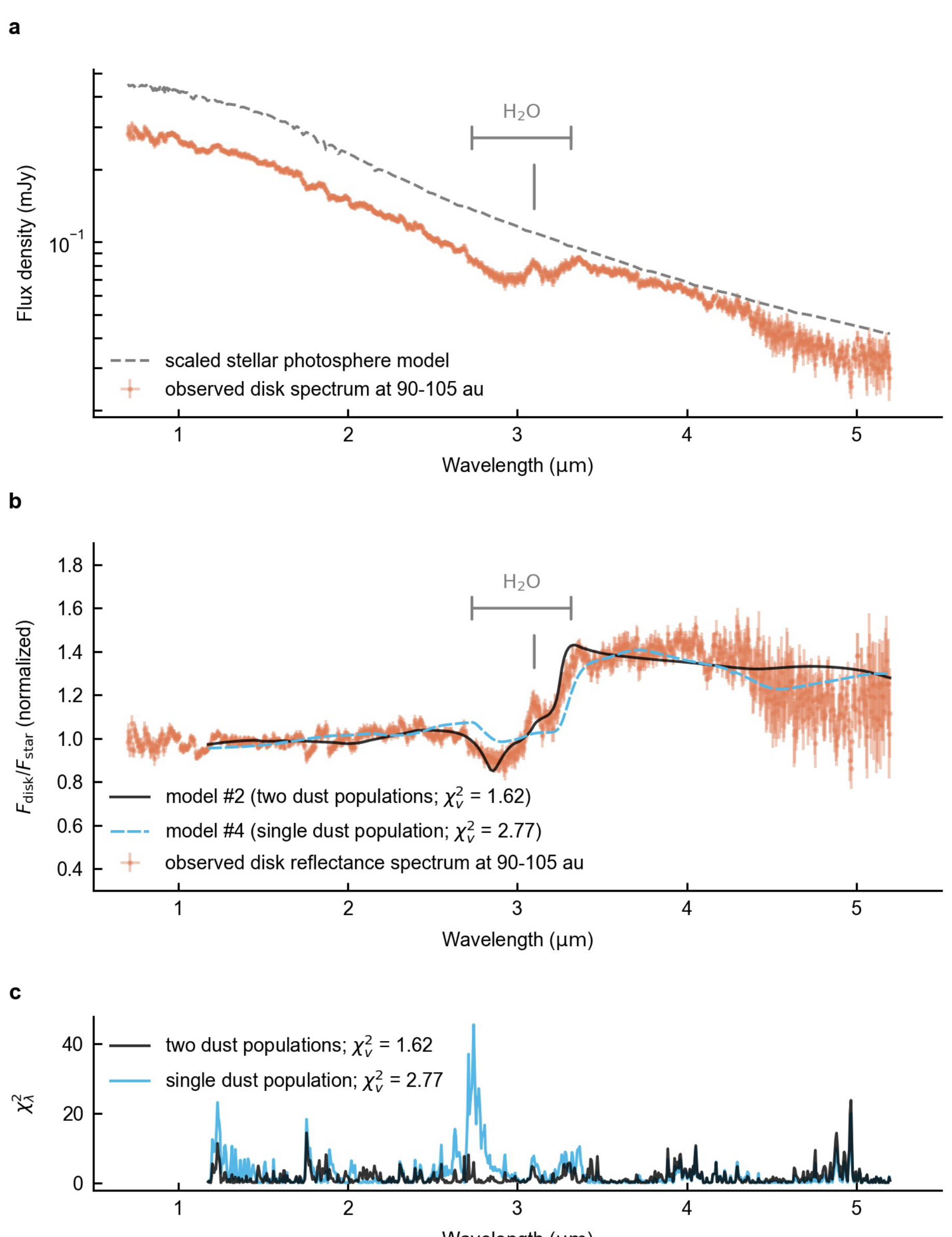

**Extended Data Fig. 3 | Illustration of reflectance spectroscopy of the HD181327 disk. a**, Comparison between the measured disk spectrum after RDI and a scaled stellar photosphere model. The extracting region of the disk spectrum is located at a scattering angle of ~90° and the stellocentric distances of 90–105 au (region 2 in Fig. 1b). The solid-state feature of water ice at 3 μm leads to changes in the slope of the disk spectrum. The Fresnel peak of water ice is presented at 3.1 μm, indicated by the vertical gray line. **b**, The disk reflectance spectrum is overlaid with best-fit dust model spectra. The solid-state feature of water ice and the Fresnel peak are visible. The bowl-shaped dip at 3 μm can be fitted using the dust model with two dust populations (model #2 in Table 1), while showing deviations when using a single dust population (model #4 in Extended Data Table 1). The Fresnel peak is not fitted because we do not include large particles (i.e., ~1 mm) in the fit. **c**, The $\chi^2$ value per spectral channel shows the performance of model fitting in each spectral channel. The dust model with a single dust population shows significant residuals (blue curve) around 2.8 μm. Error bars represent 1 s.d.

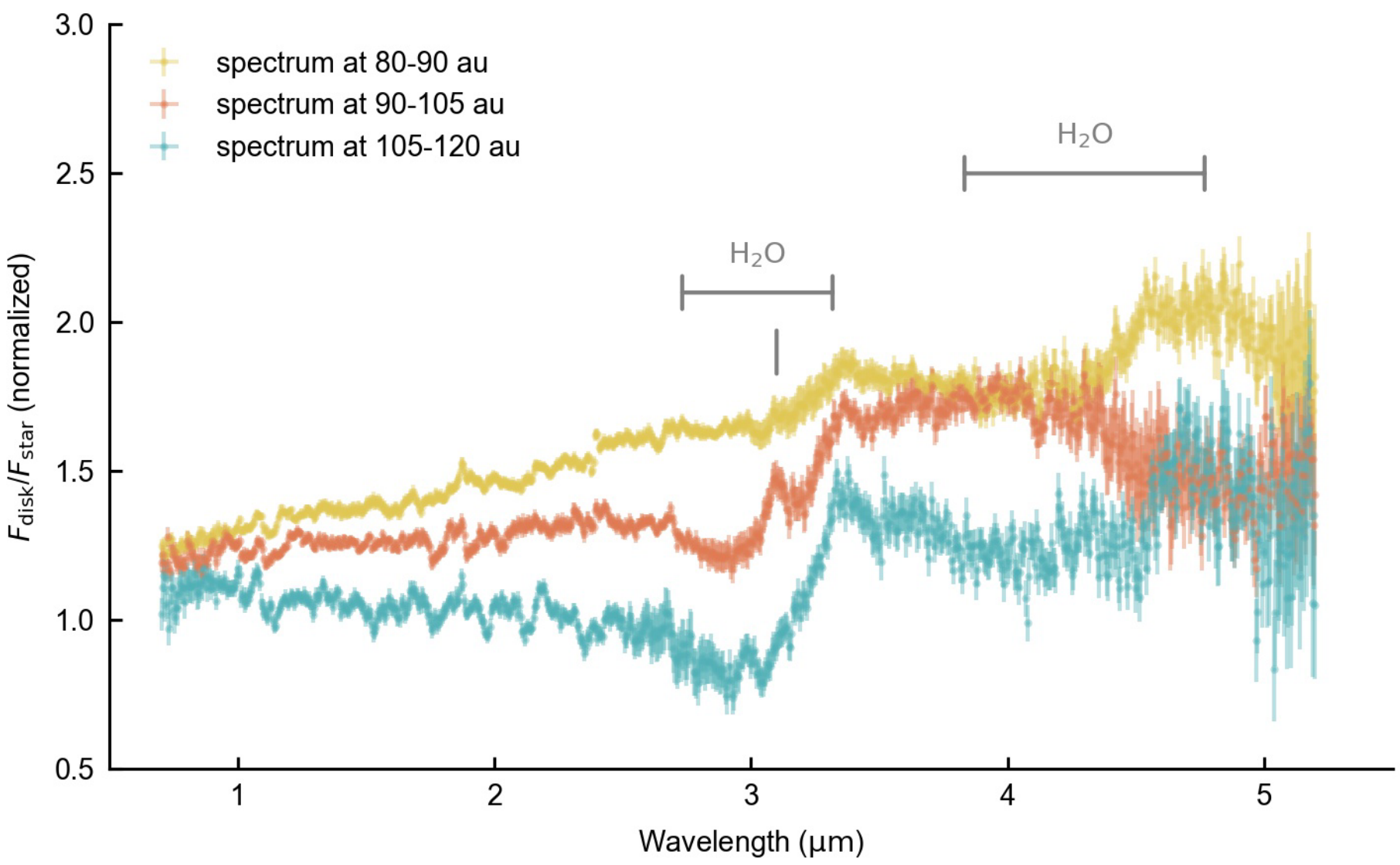

**Extended Data Fig. 4 | Disk reflectance spectra at different stellocentric distances without applying the throughput corrections.** The disk reflectance spectra at 80–90 au, 90–105 au, and 105–120 au, similar to Fig. 1, but without the throughput correction. The water-ice gradient is visible and has a similar behavior as in Fig. 1. The spectra are also bluer at larger distances, consistent with the trend found in Fig. 1. Error bars represent 1 s.d.

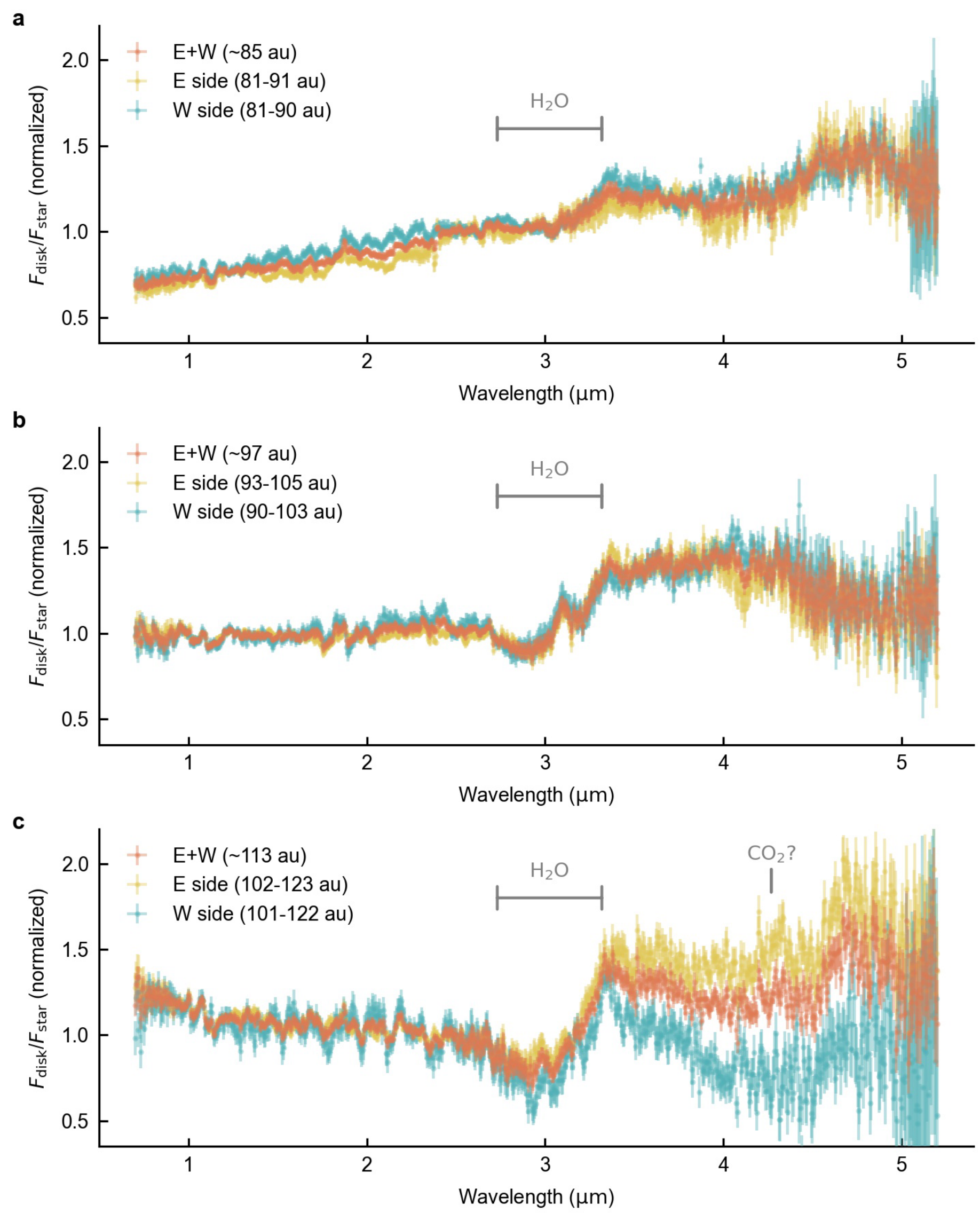

**Extended Data Fig. 5 | Disk reflectance spectra at two sides of the disk. a-c**, The disk reflectance spectra extracted from each side (east and west) of the disk at different stellocentric distances, showing no significant difference between the spectra at the two sides of the disk. The two-sides combined spectra are also shown in Fig. 1. Error bars represent 1 s.d.

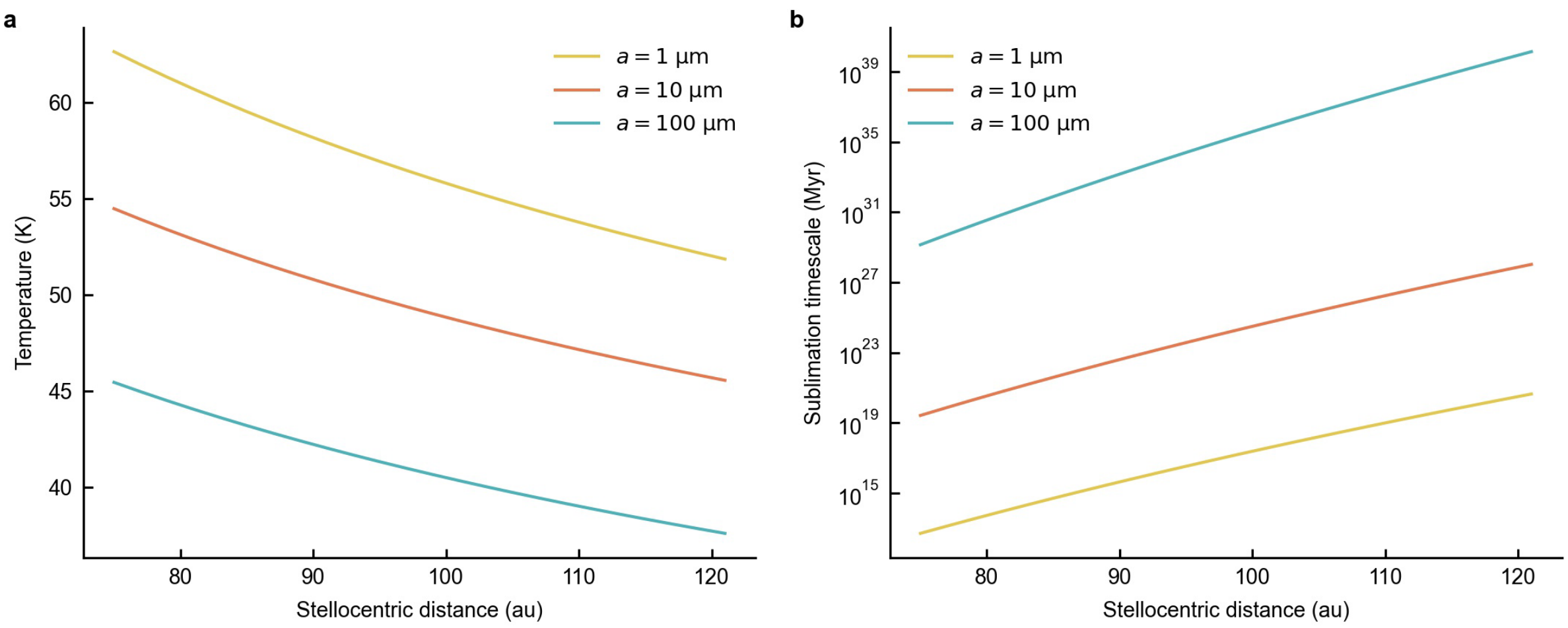

**Extended Data Fig. 6 | Dust temperatures and sublimation timescales as a function of stellocentric distance. a,** The dust temperature for different grain radii ($a$) as a function of stellocentric distance, as detailed in Methods. For micron-sized grains, the dust temperature is approximately 50 K at 80–120 au. **b**, The timescale for sublimating dirty water-ice grains of different grain sizes as a function of distance. The sublimation timescale is much larger than the stellar age of 18.5 Myr.

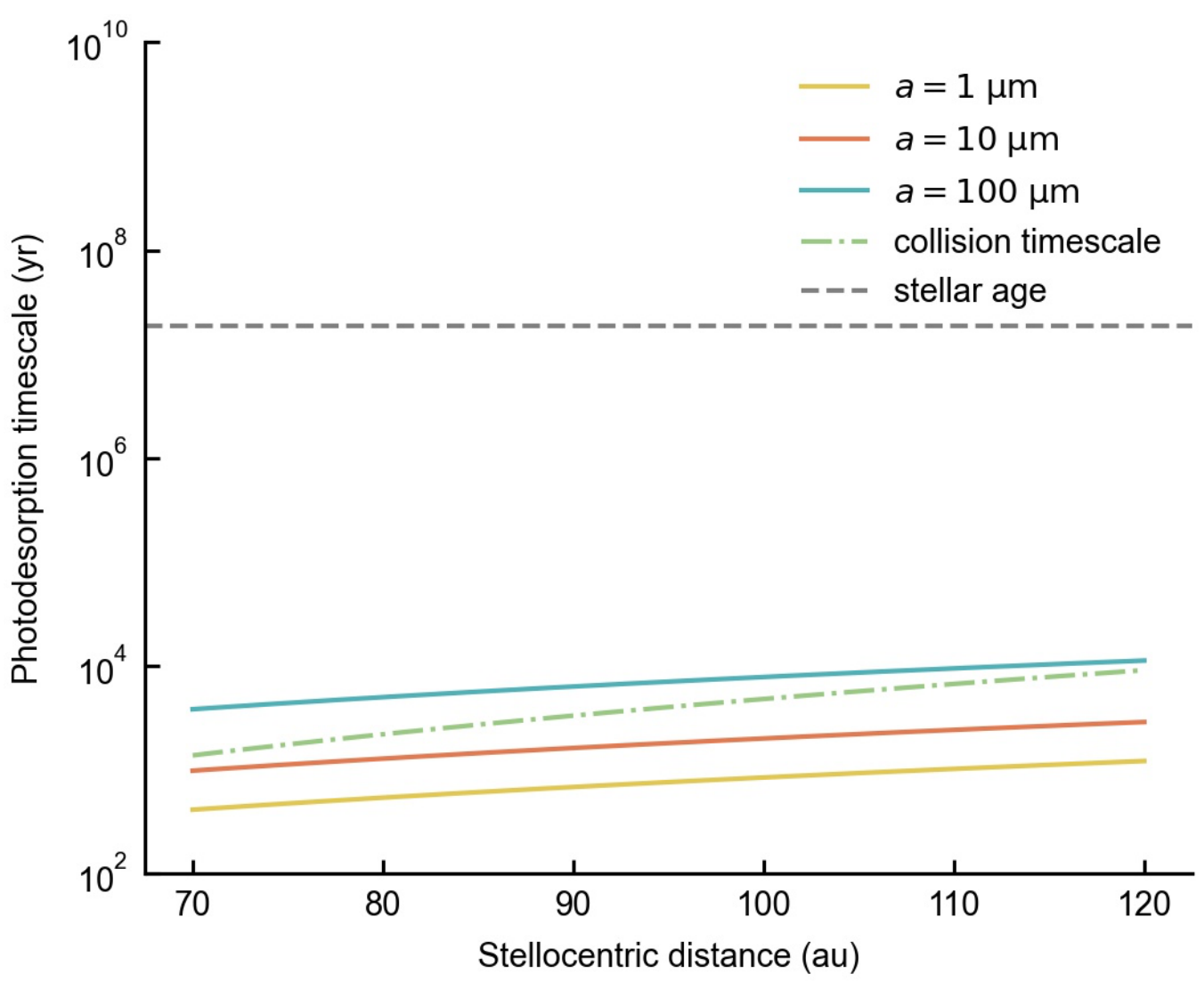

**Extended Data Fig. 7 | Photodesorption timescales as a function of stellocentric distance under the optically thin assumption.** The photodesorption timescale for destroying water ice from icy grains of different grain radii ($a$) as a function of distance, assuming optically thin ($\tau$ = 0) in the radial direction of the disk.

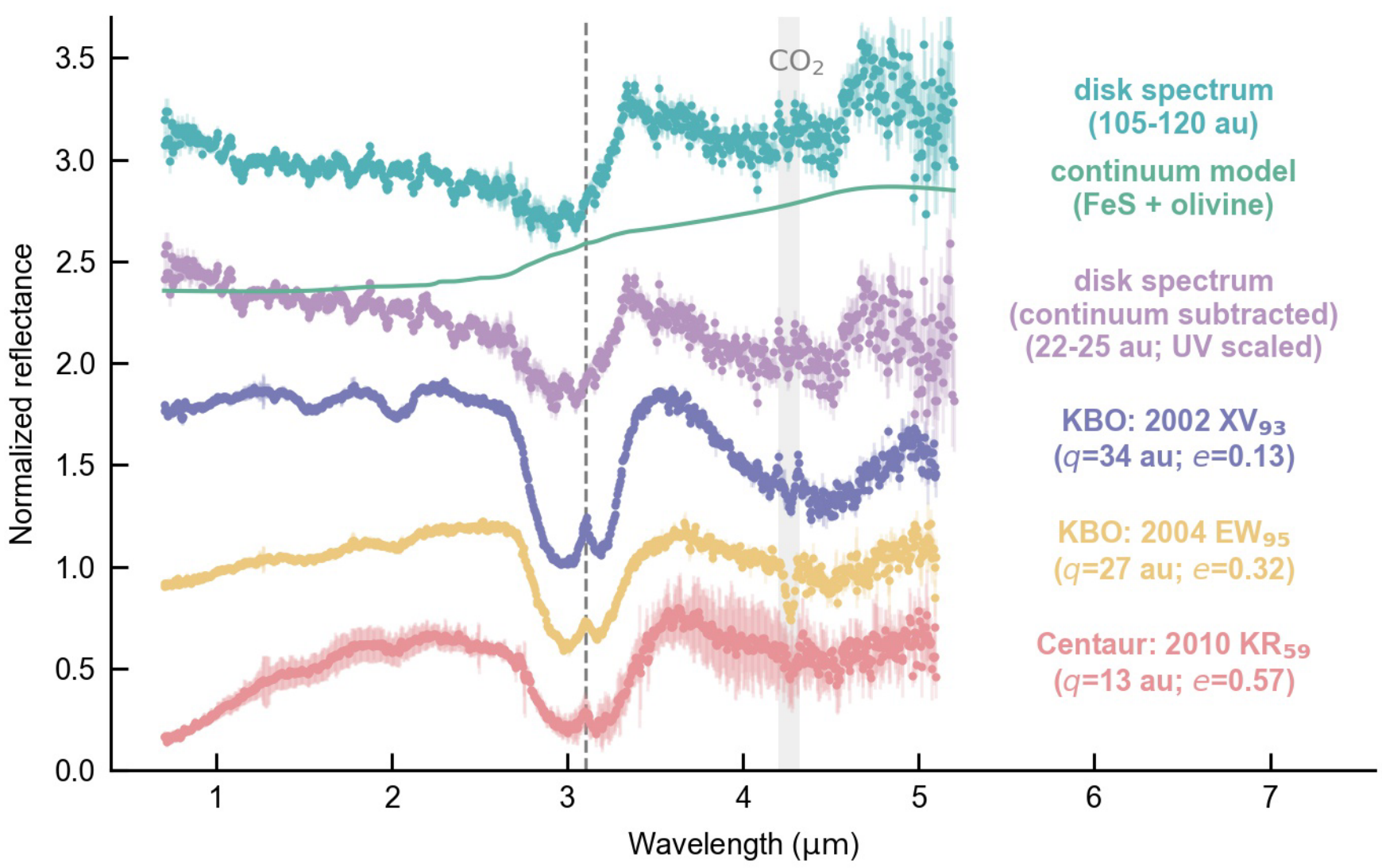

**Extended Data Fig. 8 | Comparison of the water-ice feature.** To compare the water-ice feature, a continuum model (green curve) was created and subtracted from the measured disk reflectance spectrum (cyan points). The best-fit dust parameters (model #3) from Table 1 are used to create the continuum model as detailed in Methods. The continuum subtracted spectrum (purple points) is the water-ice-dominated disk spectrum shown in Fig. 1, showing the water-ice feature at 3 μm. The spectra of water-ice-rich icy bodies are also shown for comparison, adopted from refs. [17,36], with the eccentricity ($e$) and the perihelion distance ($q$) included in the label. Although the band depths of water ice are different potentially caused by different grain

sizes and water-ice abundance, the spectra of icy bodies and the disk at locations with similar UV irradiation show a similarity in water ice. The location of the Fresnel peak at 3.1μm is marked by the vertical dashed line. Error bars represent 1 s.d.

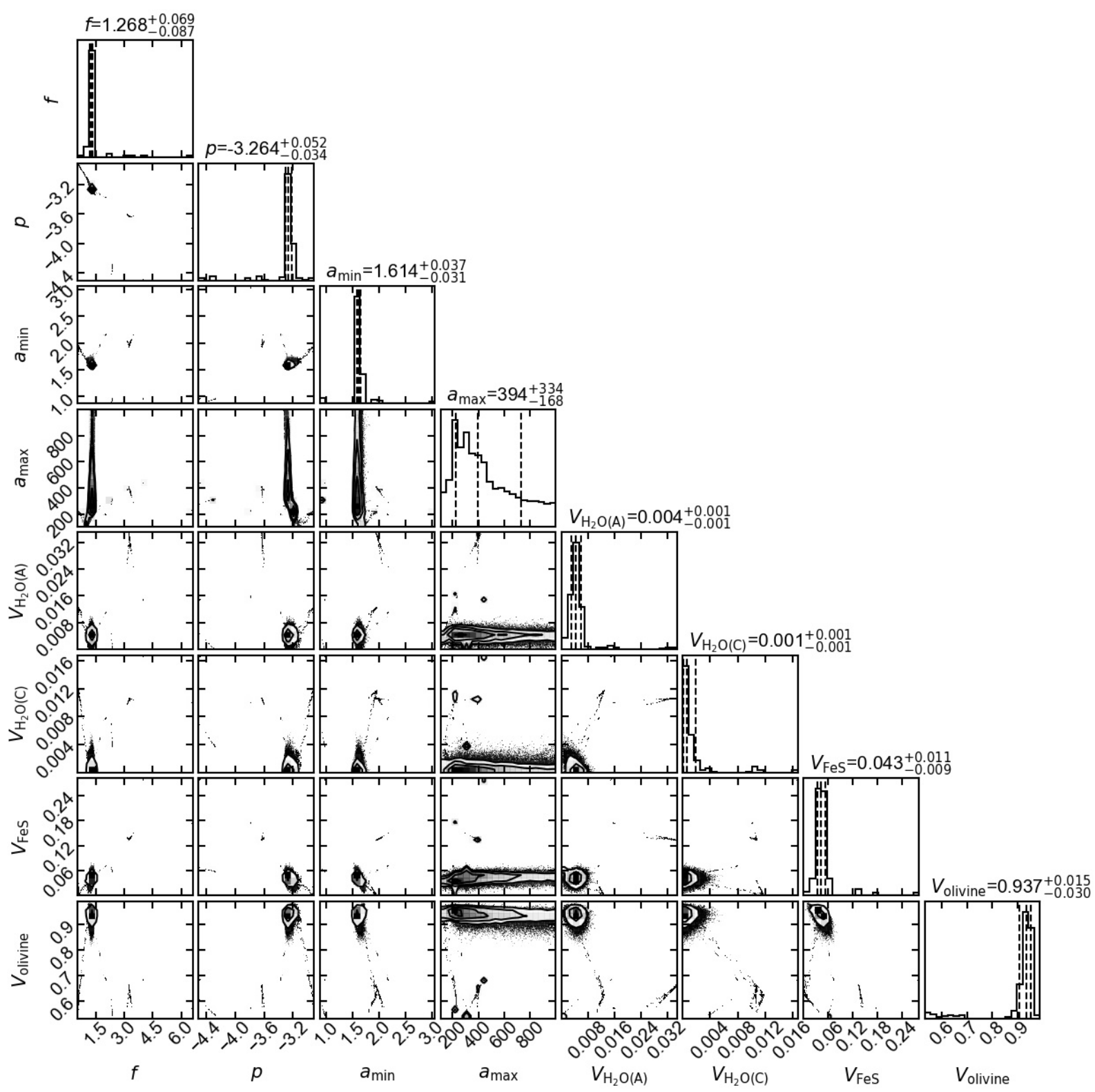

**Extended Data Fig. 9 | Corner plot of the dust parameters in model #1 at 80–90 au.** Posteriors of the dust parameters in the dust model as detailed in Methods and listed in Table 1. The vertical dashed lines denote the 16%, 50%, and 84% quantiles (68% uncertainties) of the distribution. The porosity is determined based on the volume fractions of the dust species (Methods).

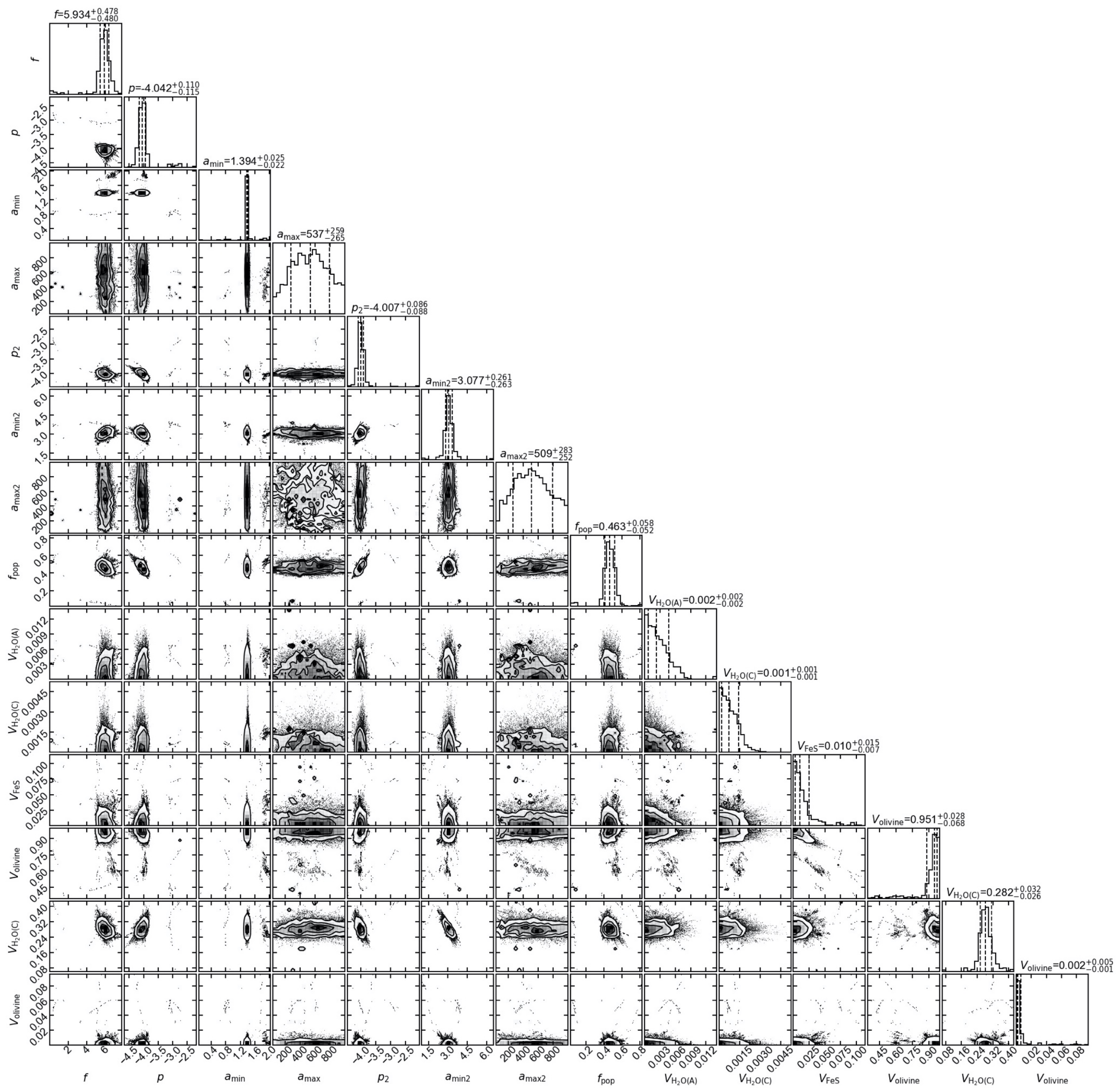

**Extended Data Fig. 10 | Corner plot of the dust parameters in model #2 at 90–105 au.** Similar to Extended Data Fig. 9. Two dust populations were adopted as listed in Table 1. The last two parameters are the volume fractions of the second dust population (i.e., the $H_2O$ and olivine mixture).

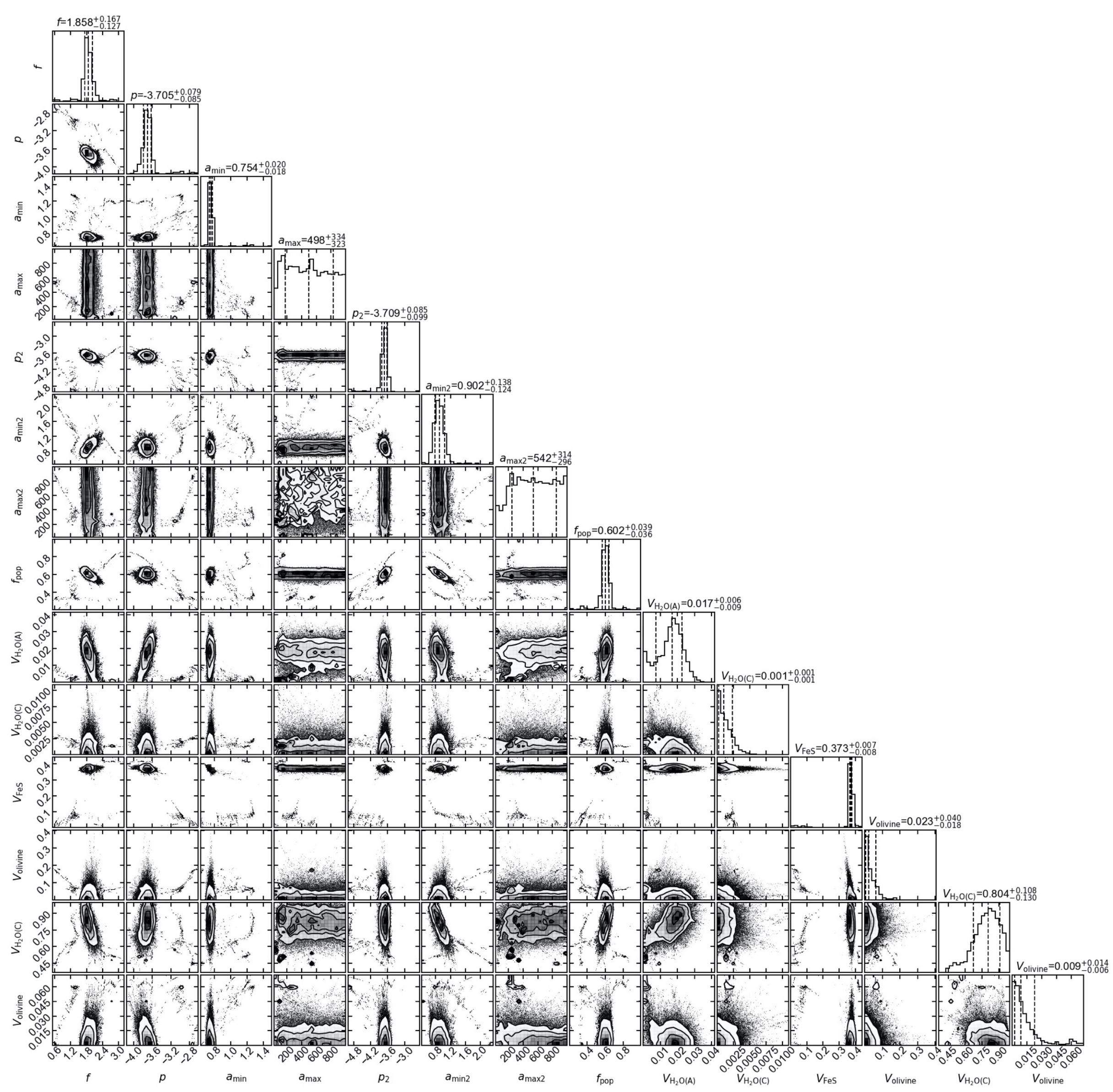

**Extended Data Fig. 11 | Corner plot of the dust parameters in model #3 at 105–120 au.** Similar to Extended Data Fig. 10.

**Extended Data Table 1 | Best-fit dust grain parameters for the model using the single dust population**

| Dust properties | Outer ring |
|---|---|
| Distance (au) | 90–105 |
| Model | #4 |
| Dust population #1 | |
| $f_{\rm pop1}$ | 1 |
| $a_{\rm min}$ (μm) | $1.35^{+0.02}_{-0.03}$ |
| $a_{\rm max}$ (μm) | $519^{+323}_{-285}$ |
| $p$ | $-3.77^{+0.03}_{-0.03}$ |
| $V_{\rm porosity}$ | $0.007^{+0.010}_{-0.014}$ |
| $V_{\rm H2O;\ amorphous}$ | $0.62^{+0.01}_{-0.01}$ |
| $V_{\rm H2O;\ crystalline}$ | $0.002^{+0.005}_{-0.001}$ |
| $V_{\rm FeS}$ | $0.37^{+0.01}_{-0.01}$ |
| $V_{\rm olivine}$ | $0.002^{+0.005}_{-0.001}$ |
| $\chi^2_\nu$ | 2.77 |